\RequirePackage{ifpdf}
\documentclass[letterpaper]{article}
\usepackage{jheppub}
\usepackage{xcolor}
\usepackage{amsmath}
\usepackage{epsfig}
\usepackage{caption}
\usepackage{subcaption}
\usepackage{soul}
\pdfoutput=1

\newcommand{\roughly}[1]{\mathrel{\raise.3ex\hbox{$#1$\kern-0.85em
\lower1ex\hbox{$\sim$}}}}
\newcommand{\lsim}{\roughly<}
\newcommand{\gsim}{\roughly>}

\def\nn{\nonumber}

\newcommand{\be}{\begin{equation}}
\newcommand{\bee}{\begin{equation}}
\newcommand{\ee}{\end{equation}}
\newcommand{\beea}{\begin{eqnarray}}
\newcommand{\eea}{\end{eqnarray}}
\newcommand{\bea}{\begin{eqnarray}}

\def\nott#1{\setbox0=\hbox{$#1$}                
   \dimen0=\wd0                                 
   \setbox1=\hbox{/} \dimen1=\wd1               
   \ifdim\dimen0>\dimen1                        
      \rlap{\hbox to \dimen0{\hfil/\hfil}}      
      #1                                        
   \else                                        
      \rlap{\hbox to \dimen1{\hfil$#1$\hfil}}   
      /                                         
   \fi}                                         %

\def\uxsl{\hbox{/\kern-.4000em$u$}}
\def\uxslsm{\hbox{\smaller/\kern-.5600em$u$}}
\def\pxpsl{\hbox{/\kern-.5000em$p$}}
\def\epssl{\hbox{/\kern-.5600em$\epsilon$}}
\def\delsl{\hbox{/\kern-.7000em$\nabla$}}
\def\lxpsl{\hbox{/\kern-.5600em$l$}}
\def\kxpsl{\hbox{/\kern-.5600em$k$}}
\def\qxpsl{\hbox{/\kern-.3900em$q$}}

\def\pref#1{(\ref{#1})}
\def\exd{{\rm d}}
\def\ol#1{{\overline{#1}}}

\def\cE{{\cal E}}

\def\cG{{\cal G}}

\def\cL{{\cal L}}

\def\cO{{\cal O}}

\def\cR{{\cal R}}

\def\mfa{{\mathfrak a}}

\def\mfg{{\mathfrak g}}

\def\mfs{{\mathfrak s}}

\def\mfU{{\mathfrak U}}

\def\ssB{{\scriptscriptstyle B}}

\def\ssD{{\scriptscriptstyle D}}

\def\ssL{{\scriptscriptstyle L}}

\def\ssN{{\scriptscriptstyle N}}

\def\ST{{\scriptscriptstyle ST}}

\def\QCD{{\scriptscriptstyle QCD}}

\def\tg{{\tilde g}}

\title{Axio-Chameleons: A Novel
String-Friendly\\ Multi-field Screening Mechanism}

\author[1]{Philippe Brax,}
\author[2,3,4]{C.P.~Burgess}
\author[5]{and F.~Quevedo}

\affiliation[1]{Institut de Physique Th\'eorique, Universit\'e Paris-Saclay,
CEA, CNRS, F-91191 Gif-sur-Yvette Cedex, France.
}

\affiliation[2]{Department of Physics \& Astronomy, McMaster University, 1280 Main Street West, Hamilton ON, Canada.
}
\affiliation[3]{Perimeter Institute for Theoretical Physics, 31 Caroline Street North, Waterloo ON, Canada.
}

\affiliation[4]{School of Theoretical Physics, Dublin Institute for Advanced Studies,
 10 Burlington Road, Dublin, 
Ireland}

\affiliation[5]{DAMTP, University of Cambridge, Wilberforce Road,  Cambridge, CB3 0WA, UK.}

\date{\today}

\abstract{Scalar-tensor theories with the shift symmetries required by light scalars are well-explored modifications to GR. For these, two-derivative scalar self-interactions usually dominate at low energies and interestingly compete with the two-derivative metric interactions of GR itself. Although much effort has been invested in single scalars (on grounds of simplicity) these happen to have no two-derivative interactions, requiring such models to explore higher-derivative interactions (that usually would be less important at low-energies). This suggests multiple-scalar sigma models as well-motivated candidates for finding new phenomena in tests of gravity. We identify a new multi-field screening mechanism appropriate for two light scalar fields (an axion and a Brans-Dicke style dilaton) that relies on their mutual two-derivative interactions. We show how very weak axion-matter couplings can introduce axion gradients that can reduce the apparent coupling of the Brans-Dicke scalar to macroscopic matter sources. We further identify a relaxation mechanism that allows this reduction to be amplified to a suppression by the ratio of the axion gradient's length scale to the source's radius (similar in size to the suppression found in Chameleon models). Unlike some screening mechanisms our proposal is technically natural and works deep within the regime of control of the low-energy EFT. It uses only ingredients that commonly appear in the low-energy limit of string vacua and so is likely to have wider applications to models that admit UV completions. We briefly discuss phenomenological implications and challenges for this scenario, which suggests re-examination of decay loss bounds and the value of equivalence-principle tests for different-sized objects. }

\begin{document}
\maketitle

\section{Introduction}

Two things motivate more detailed study of gravitationally coupled scalar fields that are light enough to be relevant to precision tests of gravity. One of these is the dramatic increase in things that can be tested due to the recent observation of gravitational waves (see for instance \cite{LIGO, PTA}).

The other motivation is theoretical, coming from the realization that such scalars can emerge naturally in the low energy limit of what we know about the UV completion for gravity.\footnote{Although this is sometimes argued as a consequence of the swampland program (see \cite{Swampland} for a recent review), that is not our starting point. Our own motivation is the ubiquity within string vacua of pseudo-Goldstone bosons for approximate shift symmetries: both compact `axionic' internal symmetries \cite{Axiverse} and approximate scaling symmetries \cite{UVShadows}. For a recent review see \cite{stringcosmo}. } Survival of scalars down to the very low energies relevant for testing gravity requires them to be extremely light, and this in turn usually requires some sort of shift symmetry to suppress their dependence in the scalar potential. In this case the lowest dimension interactions are two-derivative sigma-model interactions, which can be completely consistent with shift symmetries (such as when the target space is a homogeneous space).

\subsection{Occam vs Wilson}

Motivated by Occam's razor, the best-explored class of scalar-tensor theories extend General Relativity (GR) by adding only a single new scalar (see however \cite{Achucarro:2010jv, Silvestri:2013ne, Gergely:2014rna, Amendola:2014kwa, Leithes:2016xyh, Eskilt:2022zky}). This considerably simplifies the kinds of interactions that are possible and allows a reasonably general study of their possible consequences (for reviews see \cite{Joyce:2014kja, Burrage:2017qrf, Brax:2021wcv}). The single-scalar assumption has an unfortunate accidental side effect, however. As mentioned above, the low-energy dynamics of $N$ pseudo-Goldstone bosons subject to shift symmetries is dominated by two-derivative scalar self-interactions of the sigma-model form
\be
   \cL_{2-{\rm deriv}} = - \tfrac12 \, f^2 \, \sqrt{-g} \, \cG_{ab}(\phi) \, \partial_\mu \phi^a \, \partial^\mu \phi^b \,,
\ee
where $f$ is a characteristic scale and $\cG_{ab}(\phi)$ can be regarded as a metric on the $N$-dimensional target space spanned by the fields $\phi^a$. Because these involve two derivatives they can compete at low energies with the two-derivative interactions of GR itself.

Two-derivative scalar self-interactions turn out to be characterized by the curvature tensor built from the metric $\cG_{ab}$ (if the curvature tensor is zero then it is always possible to redefine the fields $\phi^a$ to ensure that $\cG_{ab} = \delta_{ab}$). In particular this means that two-derivative self-interactions always vanish when there is only one scalar. This is why discussions of general single-field shift-symmetric deviations of GR are driven to consider higher-derivative interactions, like those found in particular within Horndeski models \cite{Horndeski:1974wa}. The accidental absence of 2-derivative self-interactions for single scalar models combined with the widespread focus on single-field models raises the question of whether multiple-scalar models might contain surprises.

Indeed, from the theoretical point of view there is nothing in shift symmetries or UV physics that requires only a single field to survive to low energies, and it is not uncommon to find that if a single scalar (among the abundance of low-energy fields that parameterize deformations of string vacua) turns out to be light then the same is also true for other light scalars. Furthermore, the likelihood of finding very light scalars is higher should a low-energy mechanism be found that produces a technically natural solution to the cosmological constant problem \cite{Weinberg:1988cp, Burgess:2013ara}, since this automatically also provides a technically natural mechanism for suppressing the mass of generic gravitationally coupled scalars \cite{Albrecht:2001xt}. This is what happens in particular in recent approaches using low-energy scaling symmetries to suppress the vacuum energy \cite{YogaDE}.

\subsection{The Brans-Dicke problem}

In this paper we are particularly interested in the case where a light scalar arises as a low-energy dilaton -- {\it i.e.}~pseudo-Goldstone bosons for an accidental approximate scaling symmetry -- where the scale transformation of interest acts on Standard Model fields.\footnote{This motivation comes from exploring the model \cite{YogaDE} though our analysis here is largely independent of this.} Such a field couples to ordinary matter like a Brans-Dicke scalar \cite{Jordan, BransDicke, Dicke:1964pna, Brans}. Theorists usually regard very light and gravitationally coupled Brans-Dicke scalars as poison rather than catnip because of the strong observational constraints they must satisfy, such as from precision measurements within the solar system \cite{ScalarTensorTests, EPTests, Will:2014kxa, Pulsars, GWBounds, DoublePulsar, PTA}. In particular solar system tests constrain the post-Newtonian deviations from the metric predicted by GR -- including those due to a Brans-Dicke scalar -- to be at most of order $10^{-5}$ \cite{Cassini} (see \cite{Mariani:2023ubf} for a careful treatment of ephemerides when placing such constraints).

The existence of such strong constraints has sparked much interest in whether more intricate scalar-matter interactions might complicate their interpretation, leading to a number of `screening' proposals \cite{Khoury:2003aq, Hinterbichler:2010es, Hu:2007nk} (for reviews see \cite{Joyce:2014kja, Burrage:2017qrf, Brax:2021wcv}) that suppress the apparent strength with which a scalar field couples to macroscopic objects relative to the strength that would be inferred simply by summing the coupling to each of the constituent particles.

A desirable but difficult-to-obtain feature for these constructions is the ability to implement them within the low-energy limit of string vacua. The properties needed for screening often conflict with the properties required by the control of the weak-coupling or low-energy expansions within the relevant effective theory (see \cite{Hinterbichler:2010wu, Nastase:2013ik, Brax:2012mq, Padilla:2015wlv} for a discussion of several proposals within the context of Chameleon and Symmetron models \cite{Khoury:2003aq, Hinterbichler:2010es}).

\subsection{So what's new?}

In this paper we extend these constructions in several ways. Most importantly, our proposal relies crucially on the existence of (and two-derivative interactions amongst) {\it multiple} light scalars -- in particular an axion-dilaton pair -- such as frequently arise in supersymmetric models in general (and low-energy string vacua in particular). It is the use of multiple fields that allows us to evade conflict between the conditions required for screening and the conditions required to maintain control over the underlying weak-coupling and/or low-energy expansions used to derive the effective action. Indeed our mechanism requires only the two-derivative interactions of the axion and the dilaton and does not require a potential (or higher-derivative self-interactions) to exist at all for the dilaton whose couplings we are trying to suppress. We do propose weak new couplings between matter and the axion but only need these couplings being weak and do not require their dependence on the axion to have a `just-so' tuned functional form.

We instead exploit the `homeopathy' effect \cite{Homeopathy}, in which 2-derivative scalar self-interactions allow even extremely small axion gradients to have dramatic consequences on dilaton behaviour, extending earlier -- ultimately unsuccessful \cite{Brax:2022vlf, Lacombe:2023qfx} -- efforts to construct explicit screening mechanisms within a multifield framework. We here build a simple model that seems to do the job; systematically reducing the dilaton coupling of a bulk object relative to the sum of the couplings of its constituents. Furthermore, we identify circumstances where minimizing the system's energy can cause this reduction to be dramatic; suppressed by a chameleon-like factor of $\ell/R$ where $\ell$ is a microscopic scale and $R$ is the macroscopic object's radius.

The axion gradients we use to suppress the dilaton's coupling to matter are generated by the {\it axion}'s couplings to ordinary matter. Axion-matter couplings can be less dangerous in tests of gravity even though a light axion in principle mediates long-range forces \cite{Wilczek, GeorgiRandall, Khrip}. Having unsucessfully explored linear axion-matter couplings in \cite{Brax:2022vlf}, our starting assumption here is that the axion experiences both a vacuum and a matter-dependent scalar potential, $V(\mfa) + U(\mfa) n(x)$, where $n(x)$ is a measure of the local matter density ({\it e.g.}~of electrons or baryons, which we assume is negligible outside of any macroscopic sources). Crucially, we assume the minima of $V(\mfa)$ and $U(\mfa)$ differ from one another.

The presence of different axion minima inside and outside of matter drives the axion to develop nonzero spatial derivatives and these are often localized near an object's surface. In the presence of two-derivative dilaton-axion interactions the dilaton `sees' this axion gradient as a contribution to its potential that is localized at the body's surface. We show how its presence can lead the dilaton's gradient to be smaller outside the surface than inside -- effectively suppressing the object's effective dilaton `charge' relative to the naive sum of the dilaton couplings of its constituents.

At this point in the story there is no reason why the axion-generated change in dilaton charge should precisely cancel the source's naive dilaton charge to give a total charge that is close to zero. But the situation is different once the energy of the solution is minimized with respect to the various integration constants that label it. For some choices for the derivative axio-dilaton interactions minimizing the scalar energy turns out to drive the effective dilaton charge close to zero; suppressing it by $\ell/R$ where $\ell$ denotes the spatial width of the surface axion excursion and $R$ denotes the source's radius.

The crucial assumption that the axion potential has different minima inside and outside of bulk matter actually turns out to be true for the QCD axion, but we assume (unlike in the QCD case) that the vacuum axion potential is sufficiently small that it can be dominated by the matter-dependent potential {\it even within ordinary macroscopic matter} like the Sun or the Earth.\footnote{For the QCD axion nuclear densities are required for the matter potential to dominate the vacuum one.} Although the microscopic details of how the axion-matter potential arises are not required to describe the screening mechanism, their properties clearly matter when checking for phenomenological constraints. So we explore in \S\ref{ssec:OriginAxion} two examples of axion couplings to ordinary matter that would have this property.

The main dangers in these models are the consequences of having ordinary particle properties (like masses) depend on the spatially or temporally varying axion background and this strongly constrains the existence of such an axion potential on Earth. Exotic axion-dependent nuclear properties would also be hard to stomach within the Sun, due for example to the changes in nuclear interactions wrought on the neutron-proton mass difference \cite{Hook:2017psm}. Even a coupling only to leptons that changes the electron mass would have visible effects on spectral lines at the solar surface, and could affect opacities and hydrostatic equilibrium if the axion excursion only happens below the photosphere. We describe these challenges in \S\ref{ssec:OriginAxion} -- and identify the most hopeful parameter range -- while the rest of the paper focusses mainly on explaining how axion gradients can screen the dilaton. We find that the models to which one is led have small decay constants, which are usually regarded as being ruled out by energy loss from astrophysical bodies. We argue these bounds require reassessing in view of the environment dependence of the axion couplings. We also find that screening degrades the efficiency with which Brans-Dicke scalars evade equivalence principle tests, leading to deviations but only for sufficiently large objects. Although present constraints can be evaded improving these tests can be informative.

The screening effect we find in the two-field case has some phenomenological similarities with screening in single-field models. In the case of a constant density profile the single-field chameleon mechanism \cite{Khoury:2003aq} causes screening when a body's radius $R$ is larger than the scalar's Compton wavelength $\lambda$ within matter. Screening occurs because in this case the important coupling to the field scales with the body's surface area, leading to a scalar charge that is suppressed relative to its mass by order $\lambda/R$. For constant density profiles our two-field model also suppresses the dilaton's scalar charge (and so also its effects in tests of gravity) by a factor corresponding to the relative size of a thin shell, but in our case the shell is triggered by the sharp variation of the axion at the surface and so its size is specified by the width $\ell$ of the boundary layer over which the axion varies. This width depends on the {\it axion}'s matter couplings because it is the interaction with the resulting axion gradient that is responsible for suppressing the dilaton charge. In the cases where relaxation gives a large effect the suppression is again order $\ell/R$. In the end our two-field mechanism uses the interactions between the two fields in a crucial way and cannot be reduced to an effective one-field description.

Our examples also resemble single-field models in the types of interactions that are most effective in providing suppressions. For single-field models screening is not strong enough to accommodate the strong constraints in the solar system when exponential potentials and exponential couplings to matter \cite{Brax:2010gi,Brax:2012gr} are used, leading to the invocation of Damour-Polyakov's least coupling principle \cite{Damour:1994zq} (see however \cite{Olive:2007aj}). Better suppression is possible in some situations where these coupling functions have minima \cite{Brax:2010gi}.  We also find it difficult to find sufficient suppression using only exponential couplings, and find better results if the axio-dilaton interactions are minimized for some value $\phi_s$ for the dilaton.

We organize our presentation as follows. The next section \S\ref{sec:MultipleScalars} defines the classical field equations whose solutions are our main focus. These describe the general two-derivative couplings of light scalars and identifies the couplings to ordinary matter for which choices must be made. Then \S\ref{sec:Screening} specializes to the two-field axio-dilaton system and identifies the basic screening mechanism and how it depends on the field-dependence of the various coupling functions. \S\ref{sec:Pheno} explores several ways of obtaining the assumed form of axion potential from microscopic physics as well as the general phenomenological issues that these models face. We briefly summarize our results in \S\ref{sec:Conclusions}.

\section{Multiple scalars and matter}
\label{sec:MultipleScalars}

This section defines the system whose field equations are to be solved for the multiple-scalar systems coupled to matter and gravity.

\subsection{Action and field equations}
\label{ssec:ActionFE}

Consider a general sigma-model containing $N$ fields, $\phi^a$, with a target space metric $\cG_{ab}(\phi)$, in terms of which target-space proper distance is given by
\be \label{GenSigModMetric}
   \exd \mfs^2 = \cG_{ab}(\phi) \, \exd \phi^a \, \exd \phi^b  \,.
\ee
We imagine scalar self-interactions are governed by the sigma model based on this metric, with Einstein-frame lagrangian density\footnote{We use Weinberg's metric and curvature conventions throughout \cite{Weinberg:1972kfs}.}
\be \label{STaction}
   S_\ST = - \int \exd^4x \sqrt{-g} \left[ \frac{M_p^2}2 \cR + \frac{f^2}{2} \, \cG_{ab}(\phi) \, \partial_\mu \phi^a \, \partial^\mu \phi^b  + V(\phi) \right] \,,
\ee
for which the scalar field equations are
\be \label{ScalarFEGenForm}
   \frac{1}{\sqrt{-g}}\; \partial_\mu \Bigl( \sqrt{-g} \, \cG_{ab} \,\partial^\mu \phi^b \Bigr) - \frac12\,\partial_a \cG_{bc} \, \partial_\mu \phi^b \, \partial^\mu \phi^c - \frac{ V_a}{f^2} = \cG_{ab} \Bigl[ \Box \phi^b + \Gamma^b_{cd} \, \partial_\mu \phi^c \, \partial^\mu \phi^d \Bigr] - \frac{V_a }{f^2} = 0 \,,
\ee
with $V_a := \partial_a V$ and $\Gamma^a_{bc}(\phi)$ are the Christoffel symbols built from the metric $\cG_{ab}$. The trace-reversed Einstein equations similarly are
\be \label{EinsteinGCase}
   \cR_{\mu\nu} + \frac{f^2}{M_p^2} \; \cG_{ab} \, \partial_\mu \phi^a \, \partial_\nu \phi^b  + \frac{V}{M_p^2} \; g_{\mu\nu} = 0 \,.
\ee

\subsubsection*{Scalar-matter couplings}
\label{sec:AxionMatter}

Consider next how to extend the scalar-gravity models described above to include couplings with ordinary matter, doing so in a way motivated by \cite{YogaDE}. To this end we add both direct and indirect coupling to matter through a Jordan-frame spacetime metric, writing the action as $S = S_{\ST} + S_m$ where the scalar-tensor part is given by \pref{STaction} and the matter action has the form
\be
    S_m = \int \exd^4x \; \cL_m[\psi, \phi, \tg_{\mu\nu}]  \qquad\hbox{with} \qquad
     \tg_{\mu\nu} := A^2(\phi) \, g_{\mu\nu} \,.
\ee
Here $\psi$ collectively represents any matter fields and the second equation defines the Jordan-frame metric $\tg_{\mu\nu}$ in terms of the Einstein-frame metric  $g_{\mu\nu}$.

The presence of the matter action modifies the classical scalar field equations \pref{ScalarFEGenForm} to become
\be \label{ScalarFEGenFormMat}
    \cG_{ab} \Bigl[ \Box \phi^b + \Gamma^b_{cd} \, \partial_\mu \phi^c \, \partial^\mu \phi^d \Bigr] = \frac{1}{f^2} \left[ V_a + U_a  - T \, \frac{ \partial_a A}{A} \right]  \,.
\ee
Here
\be
     U_a := - \frac{1}{\sqrt{-g}} \, \left( \frac{\delta S_m}{\delta \phi^a} \right)_{\tg_{\mu\nu}\,{\rm fixed}} \,,
\ee
measures the `direct' scalar-matter couplings whose presence is not through the metric rescaling factor $A$.
In particular this would simply be $U_a = A^4 \, \partial_a \mfU$ in the special case $\cL_m = -  \sqrt{-\tg}\; \mfU$ (where the factor $A^4$ comes from the ratio $\sqrt{-\tilde g}/\sqrt{-g}$). The matter's Jordan-frame and Einstein-frame stress-energy are similarly defined by
\be
  \widetilde T^{\mu\nu} := \frac{2}{\sqrt{-\tg}} \, \left( \frac{\delta S_m}{\delta \tilde g_{\mu\nu}} \right)  \quad
  \hbox{and} \quad
  T^{\mu\nu} := \frac{2}{\sqrt{-g}} \left( \frac{\delta S_m}{\delta g_{\mu\nu}} \right) = A^6 \,\widetilde T^{\mu\nu} \,,
\ee
with traces defined by $T := g_{\mu\nu} T^{\mu\nu}$ and $\widetilde T := \tg_{\mu\nu} \widetilde T^{\mu\nu}$. These definitions imply $T = A^4 \, \widetilde T$ and so $\sqrt{-g} \; T = \sqrt{-\tg} \; \widetilde T$. Notice that general covariance ensures that when the matter fields $\psi$ satisfy their field equations stress-energy conservation for the matter sector becomes
\be
   \widetilde \nabla_\nu \, \widetilde T^{\mu\nu} = \frac{1}{A^4} \, U_a \, \tg^{\mu\nu} \partial_\nu \phi^a \,,
\ee
for arbitrary scalar and metric field configurations. In particular it is $\widetilde T^{\mu\nu}$ that is covariantly conserved in the absence of direct (non-metric) scalar-matter couplings.

The trace-reversed Einstein field equation \pref{EinsteinGCase} similarly is modified to become
\be \label{EinsteinFE}
     \cR_{\mu\nu} + \frac{f^2}{M_p^2} \; \cG_{ab} \, \partial_\mu \phi^a \, \partial_\nu \phi^b + \frac{1}{M_p^2} \left( T_{\mu\nu} - \frac12 \, T \, g_{\mu\nu} + V \, g_{\mu\nu} \right) = 0 \,.
\ee
Further progress requires us to make some choices for the nature of the scalar-matter couplings, and we do so here with an eye on the phenomenological issues discussed in \S\ref{sec:Pheno}.

\subsection{An axio-dilaton special case}
\label{ssec:TwoFieldCase}

For concreteness' sake this section explores the simplest nontrivial case, for which there are two scalars $\{\phi^a \} = \{ \phi, \mfa \}$. We take a target space metric for which one direction is a would-be axion, with a shift symmetry of the target space metric, and the most general metric consistent with this assumption can be written
\be \label{TwoFieldTSMetric}
   \exd \mfs^2 = f^2 \Bigl( \exd \phi^2 + W^2(\phi) \, \exd \mfa^2 \Bigr)
\ee
through an appropriate choice of field variables.\footnote{Pairing of dilatonic and axionic pseudo-Goldstone modes is fairly generic in the low-energy limit of string vacua.} The mass scale $f$ here sets the scale of the kinetic terms for these fields and so plays the role of the corresponding scalar `decay' constant.

We imagine the axion shift symmetry to be anomalous (as would be the case for the QCD axion) and so ultimately broken both by the vacuum scalar potential $V(\mfa)$ and by direct coupling to the matter action $\mfU(\mfa)$. We further assume $\phi$ to couple as would a pseudo-Goldstone boson for approximate scaling symmetries (such as those ubiquitous to string vacua \cite{UVShadows}), and so to couple to matter only through the scaling function: $A(\phi) = e^{\hat\mfg \phi}$, with no other direct coupling to matter. We finally assume any appearance of $V$ in both the dilaton and Einstein equations is small enough to be negligible for {\it e.g.}~solar-system applications.\footnote{The neglect of the gravitational influence of $V$ (and $\mfU$) is at first sight a strong assumption because scalar potentials are well-known to be generically UV sensitive and so difficult to arrange to be small. We do not require the details here for precisely how $V$ is arranged to be small in any particular model because our focus is on whether screening mechanisms can suppress matter-dilaton couplings in the regime where the dilaton $\phi$ is light enough to mediate macroscopic forces. For those interested, we regard \cite{YogaDE} as the best current guess as to how this might be accomplished.}

These assumptions make $\phi$ potentially dangerous because the absence of $V$ in the dilaton equation means it is effectively massless and so can mediate dangerous long-range forces. Indeed, the exponential form for $A(\phi)$ means that $\phi$ couples to matter precisely as does a Brans-Dicke scalar and long years of study of Brans-Dicke theories shows these forces compete dangerously with gravity unless the coupling $\mfg := \hat\mfg(M_p/f) \lsim 10^{-3}$ \cite{Will:2014kxa}. The focus of the rest of this paper to is to explore whether the axio-dilaton-matter interactions can change this conclusion, in such a way as to allow $\mfg \sim \cO(1)$ even if the dilaton remains extremely light.

To make the connection to Brans-Dicke theories more explicit it is convenient to use the freedom to rescale $\phi$ to make its kinetic term be $-\tfrac12 M_p^2 (\partial \phi)^2$ so that the gravity-scalar action in the absence of matter becomes
\be \label{ST2FieldActionCanon}
   S_\ST =  -\int \exd^4x \, \sqrt{-g}\; \Bigl\{ \tfrac{1}2 M_p^2\, \cR +  \tfrac{1}2 M_p^2  (\partial \phi)^2 + \tfrac{1}2  f^2 W^2(\phi) \, (\partial \mfa)^2  + V(\mfa) \Bigr\} \,.
\ee
This shows that regions with nonzero axion gradient look to the field $\phi$ like regions with an effective scalar potential proportional to $W^2(\phi)$. The plan is to exploit this to develop a suppression of effective matter-dilaton couplings. The matter action corresponding to the above choices is
\be
    S_m = \int \exd^4x \; \cL_m[\psi, \mfa, \tg_{\mu\nu}]  \qquad\hbox{and} \qquad
     \tg_{\mu\nu} = A^2(\phi) \, g_{\mu\nu} = e^{2 \mfg \phi} \, g_{\mu\nu} \,,
\ee
which uses\footnote{Recall that if all couplings in $A$ and $W$ are order unity before rescaling $\phi$ then the coefficient of $\phi^n$ becomes proportional to $(M_p/f)^n$ after rescaling.\label{FootnoteMpf}} $A = e^{\mfg\phi}$ where $\mfg := \hat \mfg M_p/f$.

With these choices the scalar field equations to be explored therefore become
\be \label{AxioDilatonFE}
    \nabla_\mu \Bigl( W^2 \, \partial^\mu \mfa \Bigr)  - \frac{1}{f^2}\Bigl( V_\mfa + U_\mfa \Bigr)  = 0 \quad \hbox{and} \quad
    \Box \phi - \frac{f^2}{M_p^2} W W' \; (\partial \mfa)^2 + \frac{\mfg T}{M_p^2} = 0  \,,
\ee
where primes denote differentiation with respect to $\phi$. These are to be solved together with the trace-reversed Einstein field equation
\be \label{Einstein2fFE}
   \cR_{\mu\nu} + \partial_\mu \phi \, \partial_\nu \phi +  \frac{f^2}{M_p^2} \, W^2(\phi) \, \partial_\mu \mfa \, \partial_\nu \mfa  + \frac{V}{M_p^2} \, g_{\mu\nu} + \frac{1}{M_p^2} \left( T_{\mu\nu} - \frac12 \, T \, g_{\mu\nu} \right) = 0  \,.
\ee

In the limit where $W' = 0$ the dilaton $\phi$ and axion $\mfa$ are largely independent of one another. It is the appearance of the coupling $\mfg T/M_p^2$ in the $\phi$ equation that reveals that it couples to matter and gravity as does a Brans-Dicke scalar, with $\mfg$ interpretable as the strength of the Brans-Dicke coupling relative to gravity.\footnote{For aficianados: $\mfg$ is related to the traditional Brans-Dicke parameter $\omega$ by the relation $2\mfg^2 = 1/(3+2\omega)$.} Meanwhile the axion responds independently to the driving terms $V_\mfa$ and $U_\mfa$.

\section{A screening scenario}
\label{sec:Screening}

We next build a screening scenario for dilaton-matter couplings, with the goal of seeing how choices for the functional forms of $V_\mfa$, $U_\mfa$ and $W$ can complicate the inference of how $\phi$ responds to the presence of matter (and thereby try to evade the stringent limits on Brans-Dicke scalars).

\subsection{Vacuum and matter-induced axion potential}

Although we do not here restrict ourselves to a QCD axion, we follow the QCD example and recognize that broken shift symmetry can generate both a vacuum scalar potential $V(\mfa)$ \cite{DiVecchia:1980yfw, GrillidiCortona:2015jxo} and a matter-dependent potential, $U(\mfa)\, n(x)$, that is proportional to the local matter -- {\it e.g.}~baryon or electron -- number density $n(x)$ \cite{AxionMatterPot}. Outside of matter the axion sees only $V_+ := V(\mfa)$ but inside matter sees the sum $V_-(\mfa) := V(\mfa) + U(\mfa) \, n(x)$. A key assumption we make is to assume that the minima $\mfa_\pm$ of the potentials $V_\pm(\mfa)$ differ: $\mfa_+ \neq \mfa_-$.

As described in \S\ref{ssec:OriginAxion} the assumption $\mfa_+ \neq \mfa_-$ is actually true for the vacuum and matter axion potentials actually generated by a QCD anomaly, with $n(x) = n_\ssB(x)$ and with $\mfa_+$ turning out to be a {\it maximum} of $U(\mfa)$. For the QCD axion $V \sim \Lambda_\QCD^4$ and $U \sim \Lambda_\QCD$ and so $U(\mfa) \, n_\ssB$ can only compete with $V(\mfa)$ for matter with nuclear density (see {\it e.g.}~\cite{Hook:2017psm, Zhang:2021mks} for a clever use of this observation to use neutron-star dynamics to constrain gravitationally coupled QCD axions). But the observation $\mfa_+ \neq \mfa_-$ becomes more universally important in the presence of a mechanism -- like those described in \S\ref{ssec:OriginAxion} -- that reduces the overall size of the vacuum potential $V$ relative to the matter-dependent part.

We henceforth assume $V$ is sufficiently small that the total axion potential within matter is minimized at $\mfa_- \neq \mfa_+$ even for the densities encountered in ordinary matter (like the Sun or Earth). Because the average solar and terrestrial densities are order unity in units of 1 g/cm${}^3 \simeq 4 \times 10^{-18}$ GeV${}^4$ the vacuum potential can be overwhelmed within the interior if we ask the vacuum potential to be of order $V \sim \Lambda^4$ with
\be \label{EquivVacScale}
  \Lambda \lsim  10 \;\hbox{keV} \left( \frac{U}{\Lambda_\QCD} \right)^{1/4} \,.
\ee

The axion mass in the vacuum is related to $\Lambda$ and the size of its decay constant $f$ by $m_\mfa \sim \Lambda^2/f$, and some values for $m_\mfa$ obtained in this way for several values\footnote{We entertain smaller values for $f$ than are normally considered for reasons described in more detail in \S\ref{sec:Pheno}.} of $\Lambda$ and $f$ are listed in Table \ref{TableMacAxionMass}. The axion mass within matter is given by a similar expression with $\Lambda \to \Lambda_{\rm mat}$ and $\Lambda^4_{\rm mat} \sim U\, n$, which gives $\Lambda_{\rm mat} \simeq 35$ keV $(U/\Lambda_\QCD)^{1/4}$ when $n \sim 10^{24}$/cm${}^3$ (as is typical in the Sun or Earth). This leads to axion masses, $m_{\rm in}$, inside matter that are similar to the first column of Table \ref{TableMacAxionMass}.

\begin{table}[ht]
\centering
\begin{tabular}{c||c|c|c}
  $m_a \; / \; m_a^{-1}$ & $\Lambda = 10$ keV & $\Lambda = 1$ eV & $\Lambda = 10^{-5}$ eV \\
   \hline\hline
   $f = 1$ MeV & $10$ keV / 2 nm  & $10^{-6}$ eV /  $20$ cm  &  $10^{-16}$ eV / $2 \times 10^6$ km  \\
    \hline
   $f = 1$ GeV & $10$ eV / 2 $\mu$m  & $10^{-9}$ eV /  $200$ m  &  $10^{-19}$ eV /  $10$ AU  \\
    \hline
   $f = 10^5$ GeV & $10^{-6}$ eV / 20 cm  & $10^{-14}$ eV /  $2 \times 10^4$ km  &  $10^{-24}$ eV /  $7$ pc  \\
    \hline
   $f = 10^{10}$ GeV & $10^{-11}$ eV / 20 km   &  $10^{-19}$ eV / $10$ AU & $10^{-29}$ eV / $0.7$ Mpc  \\
    \hline
    $f = 10^{15}$ GeV & $10^{-16}$ eV / $2 \times 10^6$ km   &  $10^{-24}$ eV / $7$ pc & $10^{-34}$ eV / $7 \times 10^{4}$ Mpc  \\
    \hline
    $f = 10^{18}$ GeV & $10^{-19}$ eV / $10$ AU  &  $10^{-27}$ eV / $7$ kpc & $10^{-37}$ eV / $7 \times 10^{7}$ Mpc  \\
    \hline
\end{tabular}
\caption{\scriptsize Values for $m_\mfa = \Lambda^2/f$ (and $m_a^{-1}$) for benchmark values of $\Lambda$ and $f$}
\label{TableMacAxionMass}
\end{table}

Choosing $\mfa_+ \neq \mfa_-$ ensures an axion gradient exists as the axion tries to minimize its energy in response to the position-dependent potential. This gradient extends into the source (and outside of it) by an amount that depends on the relative size of the local axion Compton wavelength and the distance over which the density profile significantly changes. Our interest is in situations where the Compton wavelength is the smaller of these two scales interior to the source, in which case the axion profile adiabatically follows the minimum of the local potential. This approximation breaks down if the density falls too quickly -- as it can for instance at the surface of a solid object (like the Earth) -- in which case the exterior axion field only approaches $\mfa_+$ outside the source over distances of order the external Compton wavelength. The adiabatic limit inside the source implies in particular $m_{\rm in} R \gg 1$; for comparison, the solar and terrestrial radii are $R_\odot \sim 7 \times 10^5$ km and $R_\oplus \sim 6 \times 10^3$ km respectively. The numbers in Table \ref{TableMacAxionMass} show that the condition $m_{\rm in}R \gg 1$ favours larger values of $\Lambda$ and smaller values of $f$ (see \S\ref{ssec:OriginAxion} for a discussion of the phenomenological constraints on $f$).

\begin{figure}[h]
\begin{center}
\includegraphics[width=70mm,height=40mm]{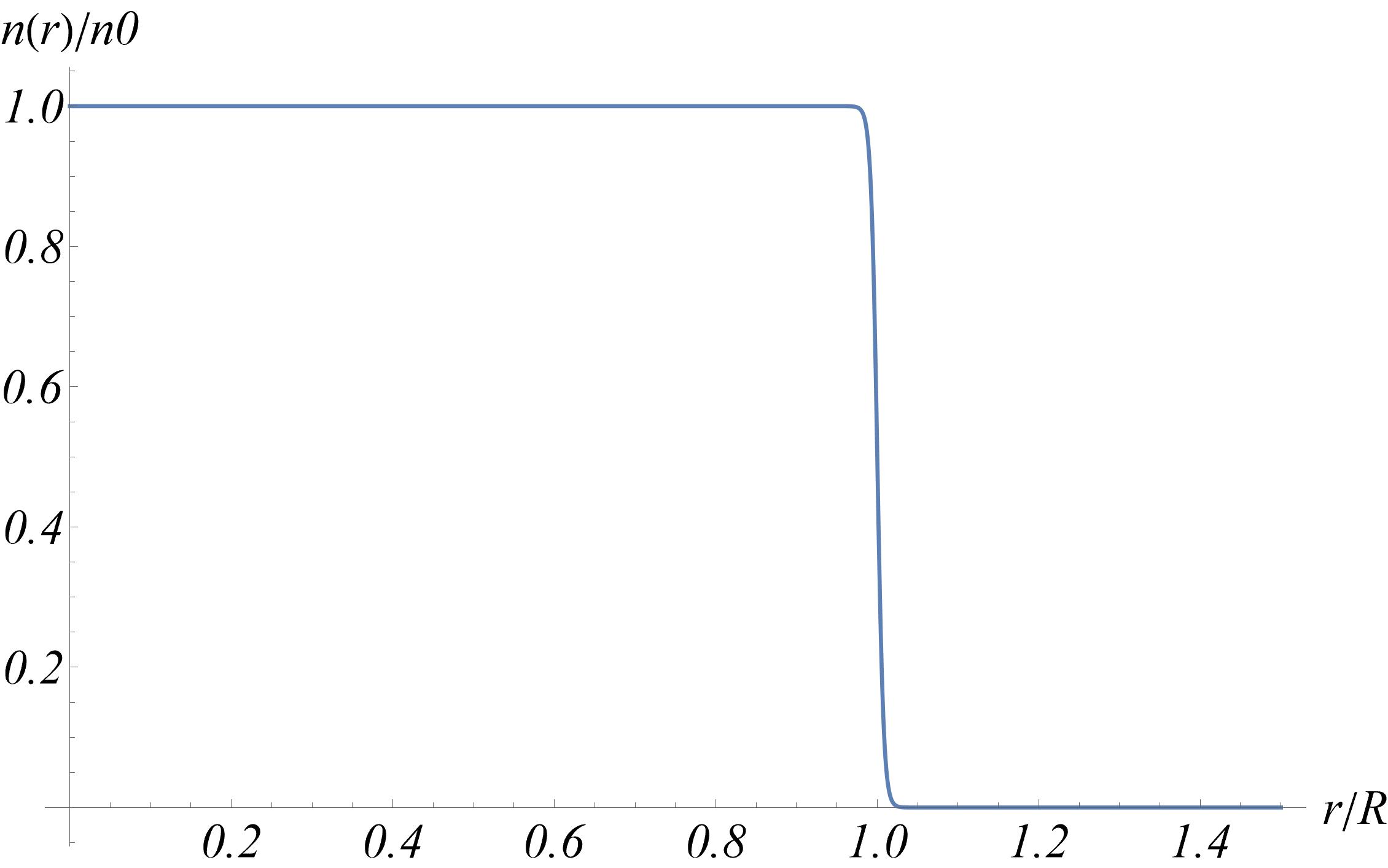}
\includegraphics[width=70mm,height=40mm]{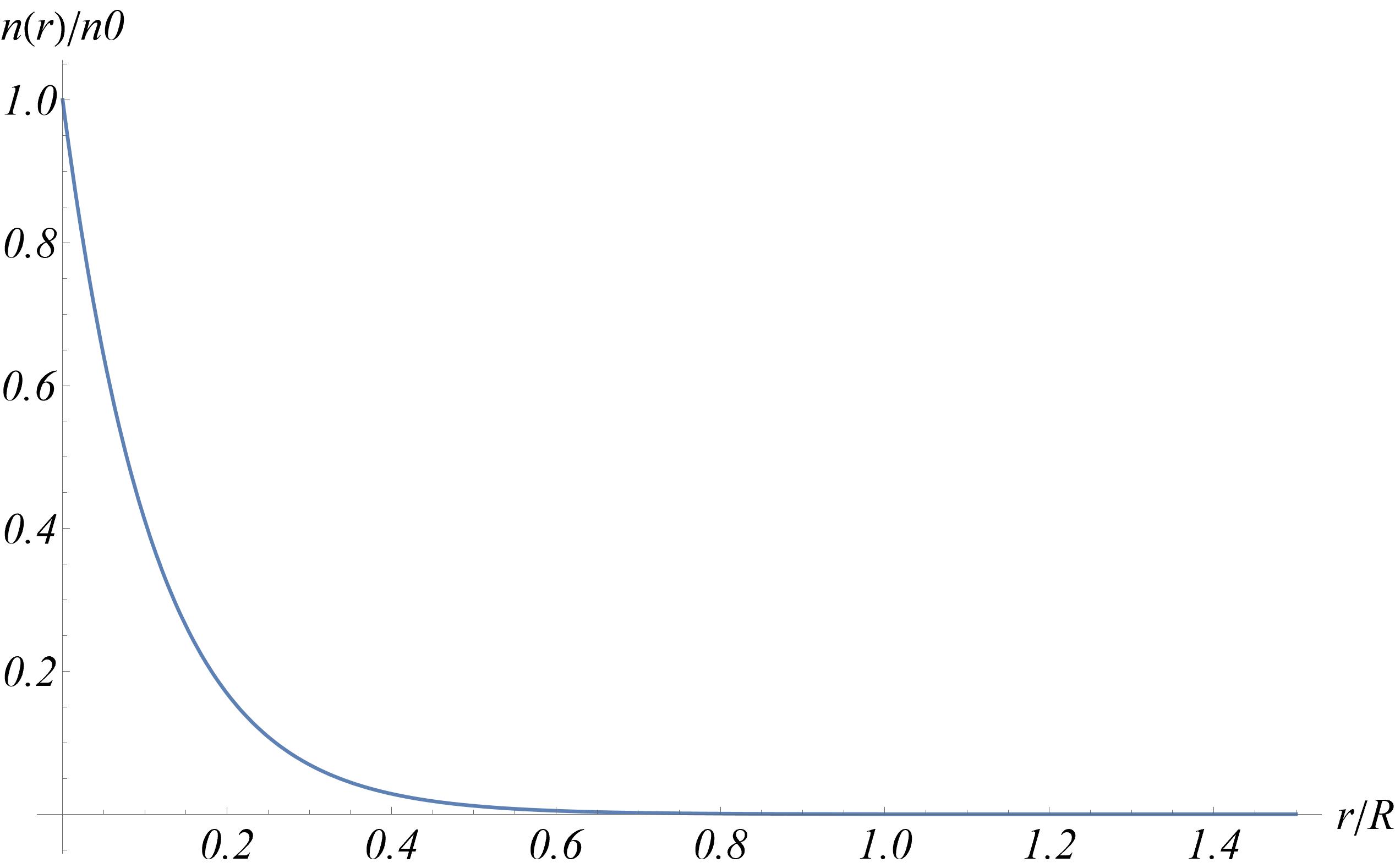}
\caption{\scriptsize Sample density profiles for use computing matter-dependent potentials. The left panel is a step function (a cartoon of the Earth) and the right panel is an exponential profile with scale-height chosen to reproduce the mean solar density.} \label{Fig:nvsrPlots}
\end{center}
\end{figure}

The width of the region containing a significant axion gradient depends on both the density profile $n(r)$ and the relative size of the axion mass, $m_{\rm in}$ and $m_{\rm out}$, interior and exterior to the source. In later sections we explore the two types of density profiles shown in Fig.~\ref{Fig:nvsrPlots}: a smoothed step function (a cartoon of the Earth's density) and an exponential function with a small step at the surface (a not-too-bad cartoon of the Sun's density). The central density and scale height of the exponential potential are chosen to agree with the central and mean solar density and the height of the step is chosen to agree with the mean density of the Earth. 

\begin{figure}[h]
\begin{center}
\includegraphics[width=70mm,height=40mm]{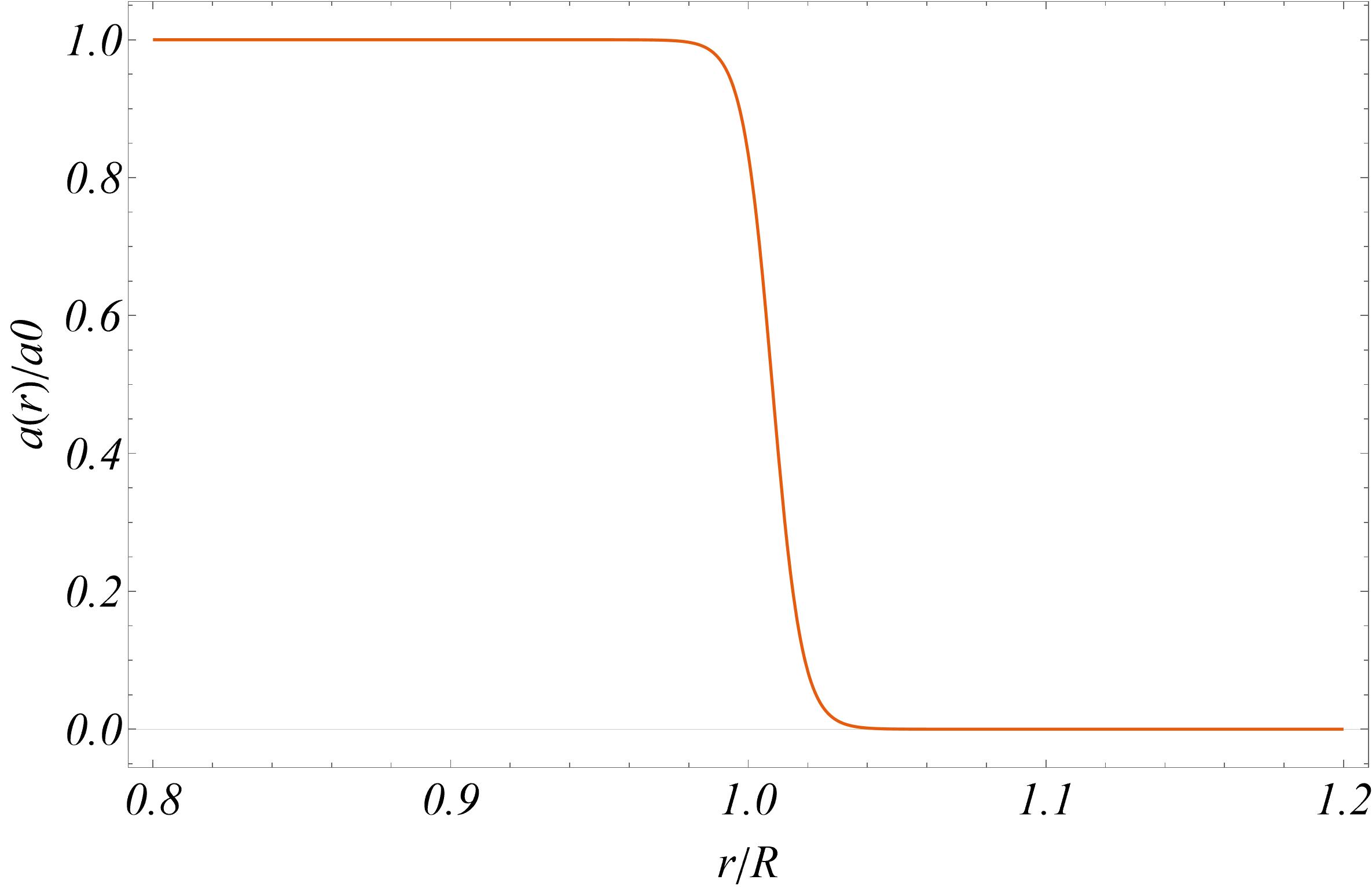}
\includegraphics[width=70mm,height=40mm]{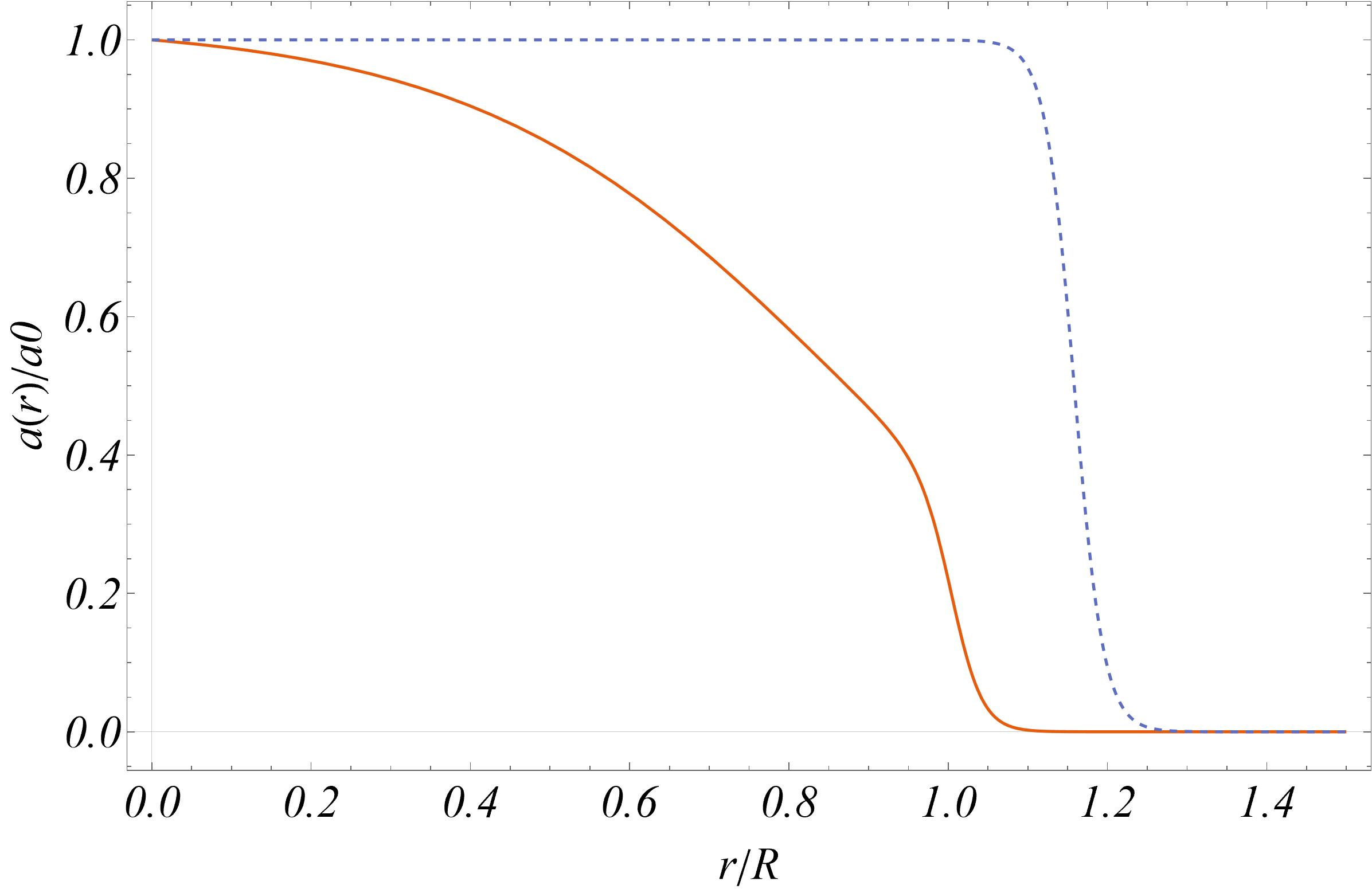}
\caption{\scriptsize Calculated axion profile $\mfa(r)$ as a function of radius in the adiabatic approximation, for the two density profiles shown in Fig.~\ref{Fig:nvsrPlots}. The external minimum is chosen to be $\mfa_+ = 0$. Left panel uses the step function profile and right panel uses the exponential profile. The solid and dashed lines in the right-hand panel differ only in the value used for the external axion mass, with the solid line using a larger value.} \label{Fig:axvsrPlots}
\end{center}
\end{figure}

Not surprisingly the axion gradient for the step-function density profile is localized near the object's surface. For the exponential profile the width of the region with nonzero axion gradient can be much larger, but becomes localized near the surface when $m_{\rm out} \ll m_{\rm in}$, as can be seen from the right-hand panel of Fig.~\ref{Fig:axvsrPlots}. The reason for this can be seen by considering the simple example $V(\mfa) = \frac12 m_{\rm out}^2 f^2 (\mfa - \mfa_+)^2$ and $U(\mfa) = \frac12 m_{\rm in}^2 f^2 (\mfa - \mfa_-)^2 F(r)$, with $0 \leq  F(r) := n(r)/n_0 \leq 1$, for which the axion field becomes
\be
   \mfa(r) = \frac{m_{\rm out}^2 \mfa_+ + m_{\rm in}^2 F(r) \,\mfa_-}{m_{\rm out}^2 + m_{\rm in}^2 F(r)}  
\ee
in the adiabatic limit. In this case $\mfa$ smoothly moves from $\mfa_+$ towards $\mfa_-$ as $F(r)$ climbs from zero as $r$ decreases from the surface, reaching $(m_{\rm out}^2 \mfa_+ + m_{\rm in}^2 \mfa_-)/(m_{\rm out}^2 + m_{\rm in}^2)$ once $r \to 0$ (since $F(0) = 1$). The $r = 0$ result closely approximates $\mfa_-$ only if $m_{\rm in}^2 \gg m_{\rm out}^2$, with $\mfa - \mfa_- = \epsilon(\mfa_+ - \mfa_-)$ for $0 < \epsilon < 1$ occuring when $r = r_\epsilon$ where $r_\epsilon$ solves
\be
   F(r_\epsilon) = \left( \frac{m_{\rm out}^2}{m_{\rm in}^2} \right) \frac{1- \epsilon}{\epsilon} \,.
\ee
This only has solutions if the right-hand side is $\leq 1$, which for $\epsilon \ll 1$ is true only if $m_{\rm out}^2 < \epsilon \, m_{\rm in}^2$. The axion gradient is localized close to the surface when the solution is $r_\epsilon \simeq R$ (for which $F(r_\epsilon) \ll 1$). Even when $\epsilon \ll 1$ this is what occurs when $m_{\rm out}^2 \ll \epsilon \, m_{\rm in}^2$ even when $F(r)$ is not dramatically step-like. 

\subsection{Dilaton surface behaviour}

We next estimate the interaction of this axion gradient with the dilaton that arises through the $W$-dependent derivative coupling terms of \pref{ST2FieldActionCanon} to see how this changes the way the dilaton responds to macroscopic matter sources. For simplicity we do so in the adiabatic limit for which the axion Compton wavelength inside matter is much shorter than the scale over which the axion varies.

For spherically symmetric solutions the dilaton equation \pref{AxioDilatonFE} (neglecting gravitational back-reaction) is
\be\label{DilatonFEAxTNoDelta}
   M_p^2\Bigl( r^2 \phi' \Bigr)' = f^2 r^2 W W' (\mfa')^2 - r^2\mfg T  \simeq f^2 r^2 W W' (\mfa')^2 + r^2\mfg \rho  \,,
\ee
where the approximate equality uses the nonrelativistic limit to replace $T \simeq - \rho$ where $\rho$ is the source's mass density, and primes on $\phi$ denote differentiation with respect to $r$ while primes on $W$ denote differentiation with respect to $\phi$. Since $T \simeq \rho = 0$ for $r > R$ the dilaton field outside the source satisfies $(r^2 \phi'_{\rm ext})' = 0$ and so
\be \label{PhiExterior}
  \phi_{\rm ext}(r) = \phi_\infty - \frac{L}{r} \,,
\ee
for integration constants $\phi_\infty$ and $L$. In particular $(r^2 \phi'_{\rm ext})_{r=R} = L$ is the quantity to which post-Newtonian tests of gravity outside the source are sensitive.

For $0 \leq r < R$ (inside the source) the spherically symmetric boundary condition $\phi'(0) = 0$ implies we instead have
\be\label{phiPrimeInterior}
   \phi'(r) 
   \simeq  \frac{1}{M_p^2r^2} \int_0^r \exd \hat r \; \hat r^2 \left[\mfg \rho(\hat r) + f^2 W W' (\mfa')^2\right]\,,
\ee
showing how the presence of an axion gradient can reduce the strength of the field generated by matter if $W'$ is negative. Continuity of $\phi'$ as one passes from inside to outside the source shows the constant $L$ is obtained by differentiating this expression and taking the limit $r \to R$ from below.

In the usual telling of the Brans-Dicke story there is no axion gradient term and so writing the total mass as $M = 4\pi \int_0^R \exd \hat r \hat r^2 \rho(\hat r)$ and using $1/M_p^{2} = 8\pi G$ then implies
\be \label{phi'Rinside}
   \phi'(R) 
   \simeq  \frac{2\mfg GM}{R^2} \,.
\ee
and so
\be
   L = 2\mfg GM \,.
\ee
This relationship permits observational constraints on the size of $L$ to be turned into limits on $\mfg$. In particular the parameterized post-Newtonian (PPN) parameter $\gamma$ turns out to be given in terms of $\mfg$ by 
\be
   \left| \gamma - 1 \right| = \frac{4\mfg^2}{1+2\mfg^2} \,,
\ee
and so the Cassini limit $|\gamma - 1| \lsim 2.3 \times 10^{-5}$ \cite{Cassini} implies $|\mfg| \lsim 2.4 \times 10^{-3}$.

\subsubsection*{Narrow-width approximation}

To explore the size of possible axion-induced reductions in $L$ consider the limit in which the axion makes its excursion from one minimum to the other within a region $R-\ell < r < R$ where $\ell \ll R$ is a small distance compared with the source's size. (As we've seen above this is often what happens in the limit $m_{\rm out}^2 \ll m_{\rm in}^2$.) In this `narrow-wall' limit the axion solution is approximately
\be \label{stepsolution}
  \mfa(r) = \mfa_- + (\mfa_+ - \mfa_-) \Theta(r-R) \quad \hbox{and so} \quad
  \mfa'(r) = (\mfa_+ - \mfa_-) \, \delta(r-R) \,,
\ee
where $\Theta(x)$ is the Heaviside step function whose derivative $\delta(x) = \Theta'(x)$ is a Dirac delta function (and primes here denote differentiation with respect to the argument). To evaluate $(\mfa')^2$ in this axion configuration we need $\delta'(0)$, which we take to define the inverse skin depth\footnote{The factor of 2 is motivated by regulating the step function as the small-$\ell$ limit of $\Theta(x) = \frac12[1 + \tanh(x/\ell)]$ and differentiating explicitly.} $\delta'(0) = (2\ell)^{-1}$, leading to
\be\label{DilatonFEAxT}
   M_p^2\Bigl( r^2 \phi' \Bigr)' \simeq \frac{f^2W W'}{2\ell} R^2(\mfa_+ - \mfa_-)^2 \delta(r-R) + r^2 \mfg \rho \,,
\ee

The presence of the delta function \pref{DilatonFEAxT} replaces continuity of $\phi'$ at $r=R$ with the jump condition obtained by integrating \pref{DilatonFEAxT} over the small interval $R-\ell < r < R$:
\be \label{phi'invsout}
  \phi'_{\rm ext}(R) \simeq \phi'(R-\ell) + \frac{f^2}{M_p^2} \left( \frac{WW'}{2\ell} \right)_{r=R} (\mfa_+ - \mfa_-)^2 \,,
\ee
where the $\mfg T/M_p^2$ term only contributes subdominantly in $\ell$ (and so is dropped). We again see that it is the sign of $W'_s = W'[\phi(R)]$ at $r=R$ that controls whether $\phi'$ is larger or smaller on the outside than on the inside of the source, with negative $W'$ ensuring $\phi'_{\rm ext}(R) < \phi'(R-\ell)$ and so reducing the dilaton `charge' of a source. Negative $W'$ is easy to arrange. For example in the example of \cite{YogaDE}
\be
  W = \frac{1}{\alpha} \; e^{- \alpha \phi} \,,
\ee
with positive $\alpha$, and so $W' < 0$ for all $\phi$.

\subsection{Relaxation mechanism}

Although it is nice that nonzero axion gradients near the surface can make the dilaton derivative $\phi'$ smaller outside a source than inside, in order to satisfy solar system tests of gravity a proper screening mechanism should be able to make $\phi'_{\rm ext}(R)$ {\it much} smaller than $\phi'(R-\ell)$. At the level of the discussion to this point this seems to be a stretch because it requires a very precise cancellation between the two terms on the right-hand side of \pref{phi'invsout}. This seems unlikely because $\phi'(R-\ell)  \propto \mfg GM/R^2$ depends on the properties of the source while the size of the second term of \pref{phi'invsout} depends on independent scales like $\ell$.

We now describe a relaxation mechanism that can accomplish this cancellation automatically. The idea starts with the recognition that $W[\phi(r)]$ evaluated at $r = R$ depends on the integration constants of the $\phi$ equation, and so in particular depends on $\phi_\infty$ of \pref{PhiExterior} -- or equivalently on $\phi_0 = \phi(r = 0)$ (which is related to $\phi_\infty$ because of continuity of $\phi$ at $r=R$). This is normally a free parameter  in the absence of a dilaton potential. Usually this integration constant is chosen to match to whatever ambient value for $\phi$ is encountered as one matches onto the rest of the universe at spatial infinity, but we instead choose to minimize the energy of the source with respect to $\phi_\infty$ (or $\phi_0$). In particular we ask whether the energetics of this adjustment can dynamically reduce the effective dilaton charge exterior to the source, and whether this energetics can compete with any small cosmological potential that $\phi$ might experience far from the source. We return to the general question of whether this minimization process is consistent with specifying the asymptotic value of the field in \S\ref{ssec:Frustration} below.

The energy in the dilaton field $\phi = \phi_\infty - L/r$ exterior to the source is given by
\be \label{EextEq}
   E_{\rm ext} = 4\pi M_p^2 \int_R^\infty \exd r \, r^2 \; \frac{ (\phi')^2}{2} = 2\pi M_p^2 L^2 \int_R^\infty \frac{\exd r}{r^2} = \frac{L^2}{4GR} \,.
\ee
The main observation is that the existence of an axion profile together with a nontrivial function $W(\phi)$ implies that $L$ is a function of $\phi_0$ (and so also of $\phi_\infty$), given explicitly by \pref{phiPrimeInterior} --- or, for the narrow-wall approximation, by \pref{phi'invsout}.

Now comes the main point. If $\phi_0$ or $\phi_\infty$ is free to be varied to minimize the energy then minimizing \pref{EextEq} shows this should prefer $L = 0$, possibly allowing deviations from GR to be negligible within the solar system despite $\mfg$ not being particularly small. But it is not quite this simple because the axion and dilaton gradients interior to the source also contains a $\phi_0$-dependent energy, given by
\be
   E_{\rm in} = 4 \pi \int_{0}^{R} \exd r \, r^2 \; \Bigl[\tfrac12 \, f^2 W^2 (\mfa')^2 + \tfrac12 \, M_p^2 (\phi')^2 \Bigr]   \,.
\ee
One must minimize $E_{\rm ext} + E_{\rm in}$ and see how large $L$ is once this is done.

\subsubsection*{Narrow-width approximation}

This minimization can be done fairly explicitly in the narrow-width approximation, in which case the interior energy is
\be
   E_{\rm in} \simeq \frac{\pi f^2}{\ell}R^2 (\mfa_+-\mfa_-)^2 W^2(r=R) + E_0 \,,
\ee
where $E_0$ denotes the dilaton contribution to the internal field energy, whose value doesn't matter because it is $\phi_0$-independent. The sum of these energies is given by $E_{\rm tot} = E_{\rm surf} + E_{\rm ext}$ and so
\be \label{EtotExp}
  \frac{4G E_{\rm tot}}{R} =  \frac{f^2}{M_p^2} \left(\frac{R}{2 \ell}\right) W_s^2(\mfa_+-\mfa_-)^2  +  \left[ \frac{2\mfg GM}{R} + \frac{f^2}{M_p^2}  \left( \frac{R}{2\ell} \right)W_sW'_s (\mfa_+ - \mfa_-)^2 \right]^2 + \frac{4G E_0}{R}
\ee
where the subscript `$s$' emphasizes that $W$ and $W'$ are evaluated at the source's surface: $\phi_s = \phi(R)$. This last expression also uses \pref{phi'invsout}, which in this context reads
\be \label{phi'invsout2}
  \frac{L}{R} = \frac{2\mfg GM}{R} + \frac{f^2}{M_p^2} \Bigl( WW' \Bigr)_{r=R} \left( \frac{R}{2\ell} \right)  (\mfa_+ - \mfa_-)^2 \,,
\ee
whose $\phi_0$-dependence emerges once $WW'(\phi)$ evaluated at $\phi(R)$ is expressed using the integral of \pref{phiPrimeInterior}.

Since the integration constant enters additively, minimizing with respect to $\phi_0$ (or $\phi_\infty$) is equivalent to minimizing with respect to $\phi_s$. Regarded as a funtion of $\phi_s$ \pref{EtotExp} has the structure
\be \label{Etotvsy}
 \frac{4 GE_{\rm tot}}{R} =  y+ \Bigl( B + \tfrac12  y'  \Bigr)^2 + \hbox{const}
\ee
where primes on $y$ denote differentiation with respect to $\phi_s$, while
\be \label{yABCdefs}
   y(\phi_s) := \frac{f^2}{M_p^2} \left(\frac{R}{2 \ell}\right) W_s^2  (\mfa_+-\mfa_-)^2    \qquad
   \hbox{and} \qquad  B = \frac{2\mfg GM}{R} \,.
\ee
As mentioned above, the good news is that the squared term is minimized when $y' = -2B$, and when this is true $L = 0$ (no dilaton charge). Although this means $y'$ is driven to be small -- suppressed by  $GM/R$ -- this affects the value of the derivative of $W^2$ and not the value of $W^2$ itself (which is good because $W^2$ appears in the axion kinetic term). The bad news is that the first ($y$) term fights this minimum,  pulling the solution away from $L=0$.

To get a feel for whether $L$ can be significantly suppressed for reasonable values of parameters we next explore several motivated choices for the functional form of $W(\phi)$.

\subsubsection{Exponential ansatz}

First suppose we make the guess that $y(\phi_s) \propto e^{-\xi \phi_s}$ motivated by the model of \cite{YogaDE}, for which $W(\phi) \propto e^{-\xi\phi/2}$. In this case $y' = - \xi y$ and so
\be
   \frac{4GE_{\rm tot}}{R} = y + \Bigl(B - \tfrac12 \xi  y \Bigr)^2 = \left(\frac{\xi}{2} \right)^2 \left[ y + \frac{2(-\xi B + 1)}{\xi^2}\right]^2 + \hbox{$y$-independent}
\ee
showing that the minimum is at
\be
   y_{\rm min} =   \frac{2B}{\xi} \left(1 - \frac{1}{\xi B} \right) \,,
\ee
provided $B \geq 1/\xi$ (and is inconsistent with $y\geq 0$ otherwise).
The dilaton charge (in units of the uncanceled charge $L_0 := 2\mfg GM$) at this minimum is
\be
 \frac{L}{L_0} = \frac{B - \tfrac12 \xi  y_{\rm min}}{B}
 = \frac{1}{\xi B}  =  \frac{ R}{2\xi \mfg G M }  \,,
\ee
where the last equality uses \pref{yABCdefs} for $B$.

When $|\mfg|$ is order unity the ratio $|L/L_0|$ is only small if $\xi \gg 1$ because this is what is required to compensate for the factor $R/GM$, which is large for weakly gravitating systems ($GM$ being of order the Schwarzschild radius). For instance for the Sun $GM_\odot/R_\odot \sim 10^{-6}$ and so having $|L/L_0|_\odot \lsim 10^{-3}$ requires $\xi \gsim 10^{9}$ if $\mfg \sim \cO(1)$. Ultimately the requirement for large $\xi$ comes from the requirement that $y'$ be small, which for exponentials also requires $y$ itself to be small. (A similar thing also happens for single-field Chameleon models, which are ineffective when using exponential coupling to matter and a exponentially decaying scalar potential -- unless the exponent of the coupling to matter is very large \cite{Brax:2010gi,Brax:2012gr}.) 

It also happens that $\xi$ can naturally be much greater than unity. As commented in footnote \ref{FootnoteMpf} the function $W$ should be expected to have order-unity couplings when expressed in terms of the original field $\phi$ whose kinetic term was $f^2 (\partial \phi)^2$. Our rescaling of $\phi$ to have Planck-scale kinetic term $M_p^2 (\partial \phi)^2$ implies the natural size to be expected for $\xi$ is $M_p/f$, which can be much larger than unity. We do not pursue this further here beyond observing that $\xi \gsim 10^9$ would require $f \lsim 10^9$ GeV, returning to the implications of this small a value for $f$ in \S\ref{sec:Pheno}.

\subsubsection{Quadratic ansatz}

Since suppression of $L$ involves $W^2$ having a small derivative we next consider the case where $\phi_s = \phi(R)$ lies near a local minimum of $W^2$. To this end imagine the field dependence\footnote{A similar ansatz was chosen for the coupling to matter in the case of the chameleon mechanism with an exponentially decaying potential \cite{Brax:2010gi}.}
\be
  W^2(\phi) = W^2_\star + \frac{W_1^2}2 \, \Bigl(\phi - \phi_\star \Bigr)^2 \,,
\ee
and so $y(\phi_s) = y_\star + \frac12 \, y_1 (\phi_s - \phi_\star)^2$, where
\be \label{y01defs}
   y_\star := \frac{f^2}{M_p^2} \left(\frac{R}{2 \ell}\right) W_\star^2  (\mfa_+-\mfa_-)^2   \quad \hbox{and} \quad y_1 :=  \frac{f^2}{M_p^2} \left(\frac{R}{2 \ell}\right) W_1^2   (\mfa_+-\mfa_-)^2  \,.
\ee
With these choices we have
\be
  y' = y_1 \, (\phi_s - \phi_\star) \,,
\ee
and so \pref{Etotvsy} becomes
\bea
 \frac{4 GE_{\rm tot}}{R} &=& \left[  y_\star + \frac{ y_1 }{2}(\phi_s - \phi_\star)^2 \right] + \Bigl[ B + \tfrac12 y_1\,(\phi_s - \phi_\star) \Bigr]^2 \nn\\
 &=& \Bigl(  y_\star + B^2 \Bigr) +  B \, y_1 (\phi_s - \phi_\star) + \frac{y_1}2(\phi_s-\phi_\star)^2 \Bigl(1 + \tfrac12 y_1  \Bigr)  \,.
\eea

Extremizing with respect to $\phi_0$ is the same as extremizing this with respect to $\phi_s$ and leads to
\be \label{phiminvsR}
     \phi_{\rm min} -\phi_\star =  -\frac{ B }{1 +  \tfrac12 y_1} = - \frac{ 2\mfg  GM/R }{1+ \frac14 (R/\ell)(f^2/M_p^2)W_1^2(\mfa_+-\mfa_-)^2} \,,
\ee
which when used in \pref{phi'invsout2} (and again denoting the Brans-Dicke result by $L_0/R = B$) implies
\bea
  \frac{L}{L_0} &=& \frac{B + \tfrac12 \, y'_{\rm min}}{B} = 1 + \frac{ y_1}{2B} \, (\phi_{\rm min} - \phi_\star) = \frac{1}{1 + \tfrac12 y_1} \nn\\
  &=& \frac{ 1}{1+ \frac14 (R/\ell)(f^2/M_p^2)W_1^2(\mfa_+-\mfa_-)^2}  \,.
\eea

Although the atoms within the source couple to $\phi$ with Brans-Dicke strength $\mfg$ (naively leading to $L_0 = 2\mfg GM$ for a macroscopic source) the axion profile at the surface actually makes the scalar couple to the macroscopic sources `as if' its Brans-Dicke coupling were
\be
   \mfg_{\rm eff} := \frac{L}{2GM} = \frac{ \mfg }{1+ \frac14 (R/\ell) \widehat W_1^2(\mfa_+-\mfa_-)^2}  \,,
\ee
where $\widehat W_1^2 = (f^2/M_p^2) W_1^2$ is the quantity that naturally is order unity (see footnote \ref{FootnoteMpf}) in the original variables before the rescaling $\phi \to \phi \, M_p/f$.

This agrees with $\mfg$ when $\widehat W_1 (\mfa_+ - \mfa_-) \to 0$ but can be much smaller if $\ell / R \ll \widehat W_1^2(\mfa_+-\mfa_-)^2$, as is always true in particular for the narrow-width limit ($\ell \ll R$) when $\widehat W_1(\mfa_+ - \mfa_-)$ is order unity. The dynamical adjustment of $\phi_0$ makes this possible, and is informative because the $\phi$-matter coupling and the non-minimal coupling of $\phi$ to the axion kinetic term break the $\phi$ shift symmetry (and so introduce a dependence of the energy on $\phi_0$).

\subsection{Asymptotic frustration}
\label{ssec:Frustration}

Why does it make sense to minimize the energy over $\phi_0$? If $\phi_0$ is determined this way then the value of the field at infinity is no longer a choice but instead becomes calculable in terms of other parameters. Why doesn't this make the solution inconsistent with any other ambient asymptotic fields provided by the distant environment? For instance, suppose $\phi$ were a cosmological field with a small scalar potential (relevant for cosmology but ignored here on the grounds that any cosmological potential is so small as to be irrelevant for the energies at play in the solar system). In this case $\phi$ would normally be expected to asymptotically approach this potential's minimum (removing the freedom to vary $\phi_\infty$).

Even worse, the value of $\phi_{\rm min}$ obtained by energy minimization depends explicitly on properties of the source, such as its radius $R$, and so why doesn't this impose contradictory conditions on $\phi_0$ when multiple sources are present (such as the Sun and other planets in the solar system, or other stars in the Milky Way)? How can the demands of different sources (with different values of $M$ and $R$, say) agree on their asymptotic value for $\phi$ far from all sources?

For multiple sources the lawyer's answer is simply that, strictly speaking, the above solutions cannot be expected to apply because of the assumption of spherical symmetry. But more usefully, having $\phi_0$ be subject to competing energetic conditions is an example of the physics of `frustration', where different types of dynamics place contradictory conditions on the value of a field. Such systems typically go to a compromise configuration for which no one condition is completely satisfied and so the various contributions to the dynamics remain partially frustrated, depending on the energy trade-off for each one. But crucially, it is trade-offs in {\it energy} that are normally paramount in establishing the ultimate compromise, suggesting that energy minimization is the right criterion for fixing $\phi_0$.

Eq.~\pref{phiminvsR} shows that for $\ell / R \ll f^2 W_1^2(\mfa_+-\mfa_-)^2/M_p^2$ screening predicts the deviation of $\phi_{\rm min}$ from $\phi_\star$ is given by
\be \label{phiminvsGMR}
  \phi_{\rm min} - \phi_\star \simeq -\frac{8\mfg \ell M_p^2}{f^2W_1^2(\mfa_+-\mfa_-)^2} \left( \frac{GM}{R^2} \right) \,,
\ee
and so depends on the particular combination $G M/R^2$ of source parameters. For a single source and in the absence of a vacuum scalar potential for the dilaton there is no other condition fixing $\phi_0$ and this is the end of the story. But for multiple sources (like in the solar system) the largest prediction for $\phi_{\rm min}$ comes from the source with the largest surface gravity, which in the solar system is the Sun (so $\phi_\odot \gg \phi_\oplus$, for example).

For multiple sources each source prefers a different asymptotic value for $\phi$ that they would like the field to take, and in the absence of a vacuum dilaton potential one expects that any asymptotic trade-off to depend on the energy cost of not being successful in achieving the prediction \pref{phiminvsGMR}, which is order $\mfg^2GM^2/R$ (which for the solar system is again largest for the Sun). For widely separated sources one expects each source to locally minimize its own energy and pay the price of there necessarily being a gradient in the field far from both sources in order to bring their asymptotic values together. The energy cost associated with this mismatch of asymptotic fields ({\it e.g.}~when matching the Earth's field to that of the Sun) is the gradient energy required to evolve between the two asymptotic solutions, $\phi_\odot$ and $\phi_\oplus$, over the distance $D$ between them (which would typically be of order several AU in the solar system).

Assuming for simplicity that $f W_1 (\mfa_+-\mfa_-)/M_p$ is order unity the energy density of this gradient is of order
\be
  \rho_{\rm grad} \sim 4\pi M_p^2 \left[ \frac{(\phi_{\oplus} - \phi_\odot)^2}{D^2} \right]  \sim \frac{\mfg^2}{GD^2} \left(\frac{\ell}{R_\odot} \right)^2 \left( \frac{GM_\odot}{R_\odot} \right)^2  \sim \mfg^2 \left(\frac{\ell}{D} \right)^2 \frac{GM_\odot^2}{R_\odot^4}
\ee
and so the total energy to support this gradient over a volume of size $R_\odot^2 D$ extending between the two sources is of order
\be
  E \sim \rho_{\rm grad} (R_\odot^2D) \sim  \left(\frac{\ell^2}{DR_\odot} \right) \frac{\mfg^2GM_\odot^2}{R_\odot} \,.
\ee
This seems to be small enough to be worth paying relative to the local energy cost, $\mfg^2GM^2/R$, of not minimizing the separate energies of the Earth and Sun when $D \gg R \gg \ell$ -- such as for solar system objects, for which $D \sim 1 \, \hbox{AU} \sim 500 R_\odot$. If this is what happens the screening should be at its best (within the solar system) for the Sun (as required to suppress tests of gravity) but also suggests that for dynamical systems the field $\phi$ cannot always optimize itself everywhere to cancel out dilaton effects. Clearly more detailed numerical studies of how the frustration energetics plays out for nonspherical configurations with multiple sources is warranted (and ongoing), as are efforts to establish whether the associated dynamical corrosion of screening can have observable consequences.

Similar considerations apply when estimating the energetic trade-offs between the predicted values for $\phi_\infty$ and the value $\phi_c$ that minimizes any cosmological dilaton potential $V_c(\phi)$ that might be relevant far from any sources. Because the energy density $V_c$ is so much smaller than the source's energy density it is well worth the energy cost of not minimizing $V_c$ in the vicinity of a source. In this case the presence of the potential also gives the dilaton a mass $m_\phi$ and the corresponding Compton wavelength sets the natural scale for the distance over which the dilaton will vary to reduce the frustration. An estimate of the change to the solution can be found by working with the potential $V_c \simeq \frac12 m_\phi^2 (\phi - \phi_c)^2$, in which case the exterior solution changes from \pref{PhiExterior} to
 \be \label{PhiExteriorMass}
  \phi_{\rm ext} = \phi_c - \frac{L}{r} \, e^{-m_\phi r}  \,,
\ee
where $L$ is again found by matching $\phi'(R)$ obtained from \pref{phi'Rinside} and \pref{phi'invsout} to
\be
  \phi_{\rm ext}'(R) = (1+m_\phi R) \frac{L}{R^2} \, e^{- m_\phi R} \,,
\ee
computed using \pref{PhiExteriorMass}. For cosmological potentials with scale $V_c \sim \mu^4$ the Hubble scale is given by $H^2 \sim \mu^2/M_p$ and so is the same size as the associated dilaton mass $m_\phi \sim \mu^2/M_p$. This makes the difference between \pref{PhiExterior} and \pref{PhiExteriorMass} only matter over cosmological distances (and so they are negligible from the perspective of describing energy frustration within much smaller systems).

\section{Phenomenological considerations}
\label{sec:Pheno}

In this section we explore some of the phenomenological issues that arise if the above screening mechanism is to be embedded into a realistic model, focussing on the challenges raised by the axion-matter couplings required to generate the desired matter-dependent potentials.

\subsection{Screening vs size}

The above discussion shows how dilaton charge is screened in an $R$-dependent way by the axion response. But this conclusion presupposes that the axion gradient is localized to a region $\ell \ll R$ and is derived using an assumption that assumes the axion Compton wavelength inside matter is much shorter than the scales $\ell$ and $R$. The opposite limit (where the interior Compton wavelength is larger than all other scales) is explored in detail in \cite{Brax:2022vlf} and does not show screening. We therefore expect screening should fail once $\ell$ and $R$ fall below a scale set by the internal axion mass.

\subsubsection{Consequences of not screening planetary sources}

As we shall see, the constraints on axion-matter couplings are the strongest if screening happens on Earth, where measurements (such as of atomic clocks) are usually most precise. So before launching into more detailed discussions of axion-matter phenomenology we first pause to remark on the solar-system implications of having dilaton couplings be screening only for the Sun and not for the Earth. This is a logical possibility because the central baryon and electron density of the Sun is higher than for the Earth, even though their averages are not so different. 

Screening of the Earth (or other planets) does not matter for the strongest solar system constraints like the Cassini bound \cite{Cassini}, because these probe only the Shapiro time delay for the motion of photons in the gravitational field near the Sun (and so constrain only the coupling $\mfg_\odot^2 \lsim 10^{-5}$, where $\mfg_\odot = \mfg_{\rm eff}$ is the Sun's screened dilaton coupling). Screening of planets does matter for other tests, however, such as the perihelion precession of Mercury or the Nordtvedt effect as measured by lunar laser-ranging experiments that measure the relative acceleration of the Earth and the Moon towards the Sun. 

For example, the constraints from the orbital motion of massive objects like Mercury are somewhat weaker than those coming from the Cassini probe, but their size also depends on the product $\mfg_\odot \mfg_p$ where $\mfg_p$ is the screened dilaton coupling for the orbiting massive object. $\mfg_p = \mfg$ is unsuppressed if the orbiting object is unscreened, and so the contraint $\mfg \mfg_\odot \lsim 10^{-4}$ coming from the precession of Mercury \cite{Will:2014kxa} can actually be slightly stronger than the Cassini limit because it is only suppressed by one power of $\mfg_\odot$. Similar conclusions apply to the bounds on the Nordtvedt parameter, $\eta_\ssN \lsim 5 \times 10^{-4}$, in Lunar Laser Ranging experiments \cite{Will:2014kxa, Battat:2023upl}.

If unscreened, the Earth's dilaton coupling would be $\mfg_\oplus = \mfg$ and so would not be surpressed in tests of Brans-Dicke theories performed on Earth or on satellites in orbit around the Earth. Constraints arise once devations are measured between the orbits predicted by the Jordan-frame and Einstein-frame metrics, $\tilde g_{\mu\nu}$ and $g_{\mu\nu}$, such as through the percent-level agreement between the predictions of GR and orbital precession measured by Gravity Probe B \cite{GPB}, since these can be interpreted as constraints on the post-Newtonian parameter $\gamma$ for the metric due to the Earth \cite{Will:2014kxa}, and so relatively weakly constrains $\mfg_\oplus^2$.

\subsubsection{Equivalence Principle tests}

Usually Brans-Dicke scalars are impervious to tests of the equivalence principle, which for the MICROSCOPE collaboration \cite{MICROSCOPE} are extremely good (one part in $10^{15}$) for objects in Earth orbit. Lunar Laser Ranging imposes almost as strong constraints on the equivalence principle as applied to the relative acceleration of the Earth and Moon towards the Sun. These tests are normally not dangerous for dilaton couplings because the equivalence principle is automatically satisfied whenever matter couples only through a metric like $\tilde g_{\mu\nu}$ (and this is part of the appeal of these kinds of couplings). 

The situation changes once screening occurs, however, because the underlying axion couplings distinguish between particle species and the amount of screening depends on an object's size \cite{Hees:2018fpg}. As we shall see -- see \S\ref{ssec:SimplifiedAxMat} below -- viability of axion phenomenology is easier the lighter the interior mass of the axion is, we work through the implications of an optimal benchmark choice that puts the interior Compton wavelength around 200 km. For such a choice screening only occurs for objects that are much larger than a few hundred kilometres, leading to the expectation that equivalence-principle violations should also only occur for objects larger than this. (Similar considerations apply to single-field Chameleon screening \cite{Khoury:2003aq}.)

For the choice $m_{\rm in}^{-1} \lsim 200$ km we do not expect strong bounds from the MICROSCOPE experiment because the objects whose orbital motions are studied are too small to be screened (implying they couple to the dilaton with strength $\mfg$ and obey the equivalence principle). This would not be true for the Earth, planets or the Sun, however, which is why Lunar Laser Ranging bounds on $\eta_\ssN$ are relevant.

\subsection{ALPs and their ilk}
\label{ssec:OriginAxion}

The screening mechanism of earlier sections is promising, but leaves open the key question as to whether the axion-matter couplings required to screen the dilaton could have themselves hitherto escaped detection. The success of the proposal requires an assessment of whether the required axion-matter couplings are themselves viable. This in turn requires a more precise microscopic specification of how these interactions arise. 

The remainder of this section explores this question in two steps. We first explore traditional axion (ALP) models, for two reasons. First, they contain a fairly limited class of interactions about which a great deal is already known, most notably including the phenomenological constraints that limit their strength and form. Second, standard QCD axions provide the best motivated example where axions acquire both vacuum and matter potentials and for which these do not share the same minima. We find the required couplings to be very difficult to accommodate within this framework, so we then step back and ask whether the couplings required for screening are possible within a broader, more phenomenological, approach. 

The  leading dimension-4 matter couplings of a dimensionless axion-like particle (ALP) to microscopic matter is given by
\be \label{AxEFT0}
  -\frac{\cL_{\rm ax}}{\sqrt{-g}} = \tfrac12 \hat f_a^2 (\partial \mfa)^2 + \kappa_s \left(\frac{\alpha_s}{8\pi} \right) \mfa \, G^a_{\mu\nu} \widetilde G_a^{\mu\nu} +\kappa_e \left(\frac{\alpha}{8\pi} \right)  \mfa \, F_{\mu\nu} \widetilde F^{\mu\nu} + \sum_\psi\tfrac12 C^0_\psi \, \partial_\mu \mfa\, (\ol \psi \gamma^\mu\gamma_5 \psi)  \,
\ee
where $\alpha_s$ and $\alpha$ are the QCD and electromagnetic fine-structure constants, $\psi$ is a matter fermion, $F_{\mu\nu}$ is the electromagnetic field and $G^a_{\mu\nu}$ is the gluon field strength, for which tilde's denote their duals. The constants $\kappa_s$ and $\kappa_e$ are respectively the model-dependent mixed colour and electromagnetic anomalies for the axion symmetry. The constants $C^0_\psi$ are model-dependent choices for the strength with which the axion can couple to fermion bilinears in a shift-symmetric way.

\begin{figure}[h]
\begin{center}
\includegraphics[width=120mm,height=80mm]{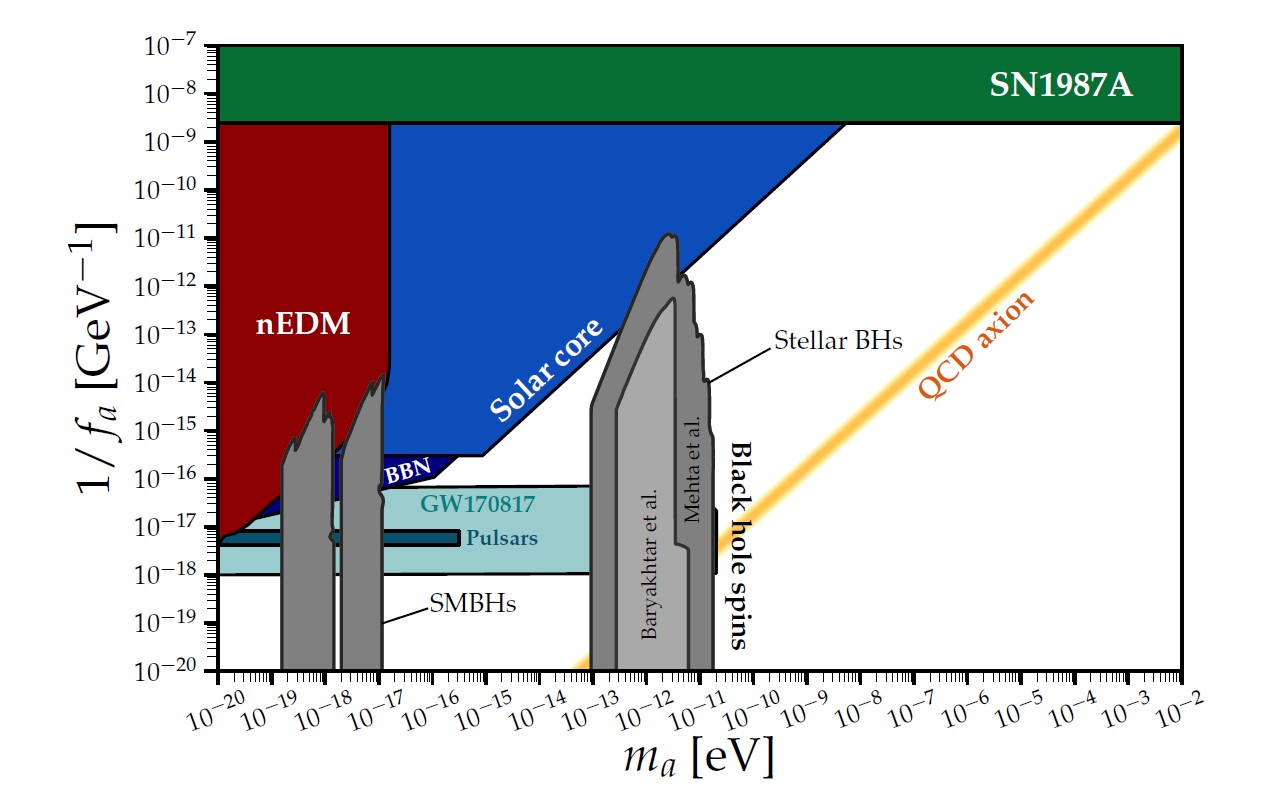}
\caption{\scriptsize Constraints on axion decay constant vs mass, showing in particular the constraints on very light Planck-coupled axions derived in \cite{SuperradianceBounds, Hook:2017psm, Zhang:2021mks}. (Figure taken from \cite{PDG, gitFigs}.)} \label{Fig:FaxBound}
\end{center}
\end{figure}

\subsubsection{QCD axion benchmark}
\label{sssec:QCDBench}

The best motivated axion is the QCD axion, defined as one for which the axion symmetry has a QCD anomaly (and so $\kappa_s \neq 0$). In this case an appropriate field redefinition allows the $G\widetilde G$ term (and its counterpart within the Standard Model lagrangian) to be rotated away, at the expense of introducing the axion into the fermion mass terms, leading to the following equivalent form
\be \label{AxEFT}
  -\frac{\cL_{\rm ax}}{\sqrt{-g}} = \tfrac12 (\partial a)^2 + \tfrac14 g_{a\gamma\gamma} \, a \, F_{\mu\nu} \widetilde F^{\mu\nu} + \sum_\psi \frac{C_\psi}{2f_a} \partial_\mu a \, (\ol \psi \gamma^\mu\gamma_5 \psi) + \cL_{\rm non-deriv} \,
\ee
where\footnote{When $\kappa_s \neq 0$ it is conventional to absorb it into $f_a$.} $a = \mfa \hat f_a = \mfa f_a \kappa_s$ is the canonically normalized axion,
\be
   g_{a\gamma\gamma} =\frac{\alpha}{2\pi f_a} \left(\frac{\kappa_e}{\kappa_s} - \Delta \right)
\ee
where $\Delta$ is a contribution that depends on the redefinition used to get from \pref{AxEFT0} to \pref{AxEFT}. For a QCD axion acting only on the two lightest quarks we have $\Delta \simeq \frac23 (4m_d+m_u)/(m_d+m_u)$ where $m_u$ and $m_d$ are the up and down quark masses. The new coupling $C_\psi$ is often combined into the dimensionless combination  $g_{a\psi\psi} := {C_\psi m_\psi}/{f_a}$. The point of this redefinition is to put all non-derivative terms in an effective axion-dependent fermion mass
\be
  -\frac{ \cL_{\rm non-deriv}}{\sqrt{-g}} = \ol \Psi M(a) \gamma_\ssL \Psi + \hbox{h.c.}
\ee
where $\Psi$ is a column matrix containing all of the fermions, $\gamma_\ssL$ is the projector onto left-handed fermions, $M(a) = \exp[\frac{i}{2} \mfa \, Q_a] M \exp[\frac{i}{2} \mfa Q_a]$ -- where $M$ is the diagonal fermion mass matrix -- and $Q_a$ is the matrix generator of the axion symmetry acting on $\Psi_\ssL$, normalized so that Tr $Q_a = 1$.

QCD generates both a vacuum and matter-dependent potential, and these are computed by evaluating the expectation value of the contribution of the energy coming from $\cL_{\rm non-deriv}$, which in the new basis can be done simply by tracking how this energy depends on the fermion masses. In vacuum if the axion symmetry acts only on the lightest two quarks then the potential becomes \cite{DiVecchia:1980yfw, GrillidiCortona:2015jxo}
\be \label{QCDVpred}
    V_\QCD(\mfa) = \Lambda^4 v(\mfa) \quad \hbox{with} \quad
    v(\mfa) \simeq -   \left[ 1 - \frac{4m_u m_d}{(m_u + m_d)^2} \sin^2 \left( \frac{\mfa}{2} \right)  \right]^{1/2} \,,
\ee
where $\Lambda \sim \Lambda_\QCD$. Famously $u(-\mfa) = u(\mfa)$ (which gets corrected at loop level only by very small CP-violating corrections) and $u(\mfa)$ is minimized at $\mfa = 0$ (up to periodic shifts). 

The matter potential similarly involves the baryon density, $n_\ssB(x)$, because the leading mass-dependence in the energy density of a macroscopic body is $\rho(x) \simeq m_\ssN n_\ssB(x)$, where $m_\ssN$ is the nucleon mass. This leads -- in the limit of equal quark masses $m_u \simeq m_d \simeq m_q$ -- to the matter potential \cite{AxionMatterPot}
\be \label{QCDmattax}
  U_{\rm ax-mat}
  =  \sigma_\ssB \, n_\ssB \, z(\mfa) \quad \hbox{with} \quad
  z(\mfa) \simeq \left| \cos \left( \frac{\mfa}{2} \right)  \right|
\ee
where $\sigma_\ssB \simeq \sum_q m_q (\partial m_\ssN/\partial m_q) \simeq 59$ MeV measures the dependence of the nucleon mass $m_\ssN$ on the quark masses. As advertised, the minimum of \pref{QCDVpred} ({\it e.g.} $\mfa = 0$) is a local {\it maximum} of \pref{QCDmattax}, though for QCD this only alters the minimum inside matter for nuclear densities (for which $n_\ssB \sim \Lambda_\QCD^3$). For densities smaller than this the axion potential is dominated by the vacuum contribution both inside and outside of matter.

\begin{figure}[t]
\begin{center}
\includegraphics[width=120mm,height=80mm]{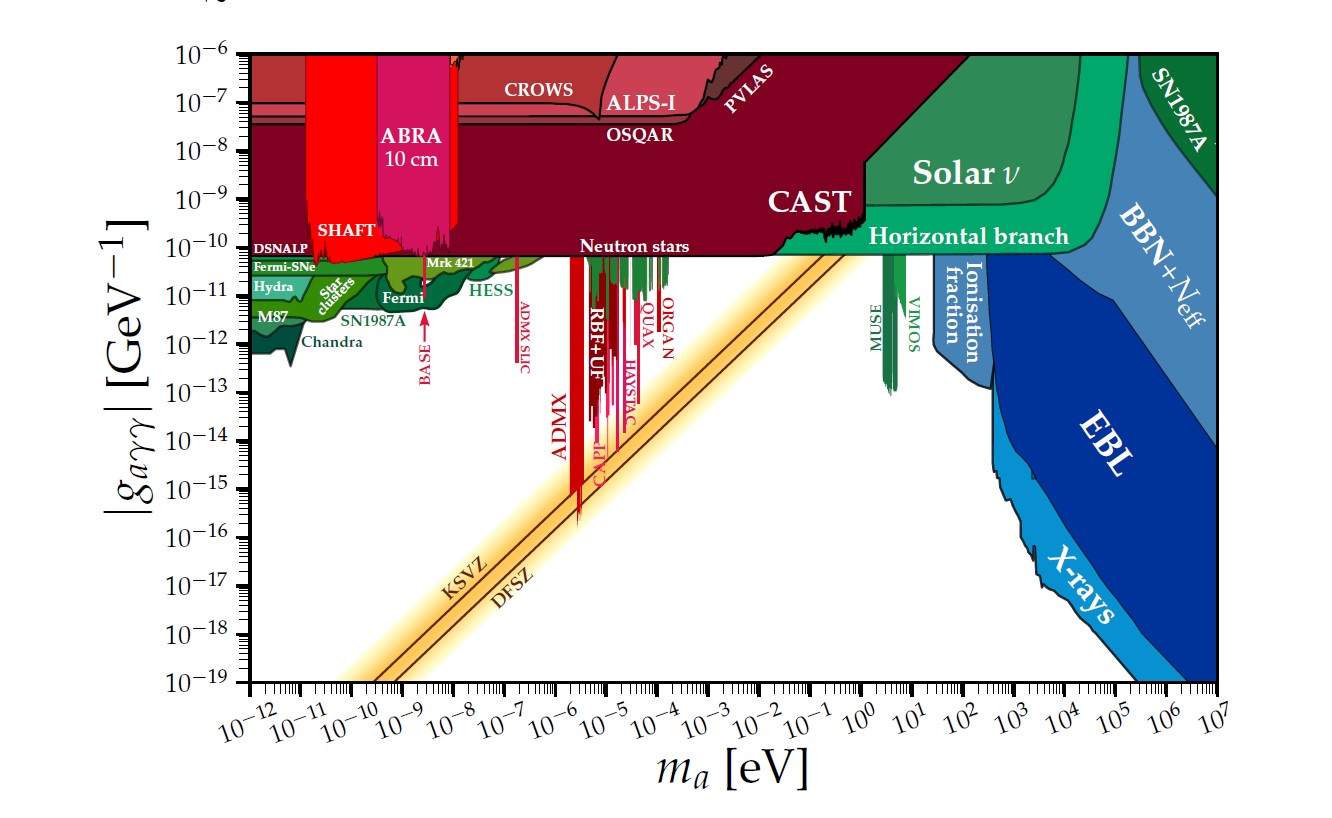}
\caption{\scriptsize Constraints on axion-photon coupling vs mass. (Figure taken from \cite{PDG, gitFigs}.)} \label{Fig:FapBound}
\end{center}
\end{figure}

\begin{figure}[h]
\begin{center}
\includegraphics[width=120mm,height=80mm]{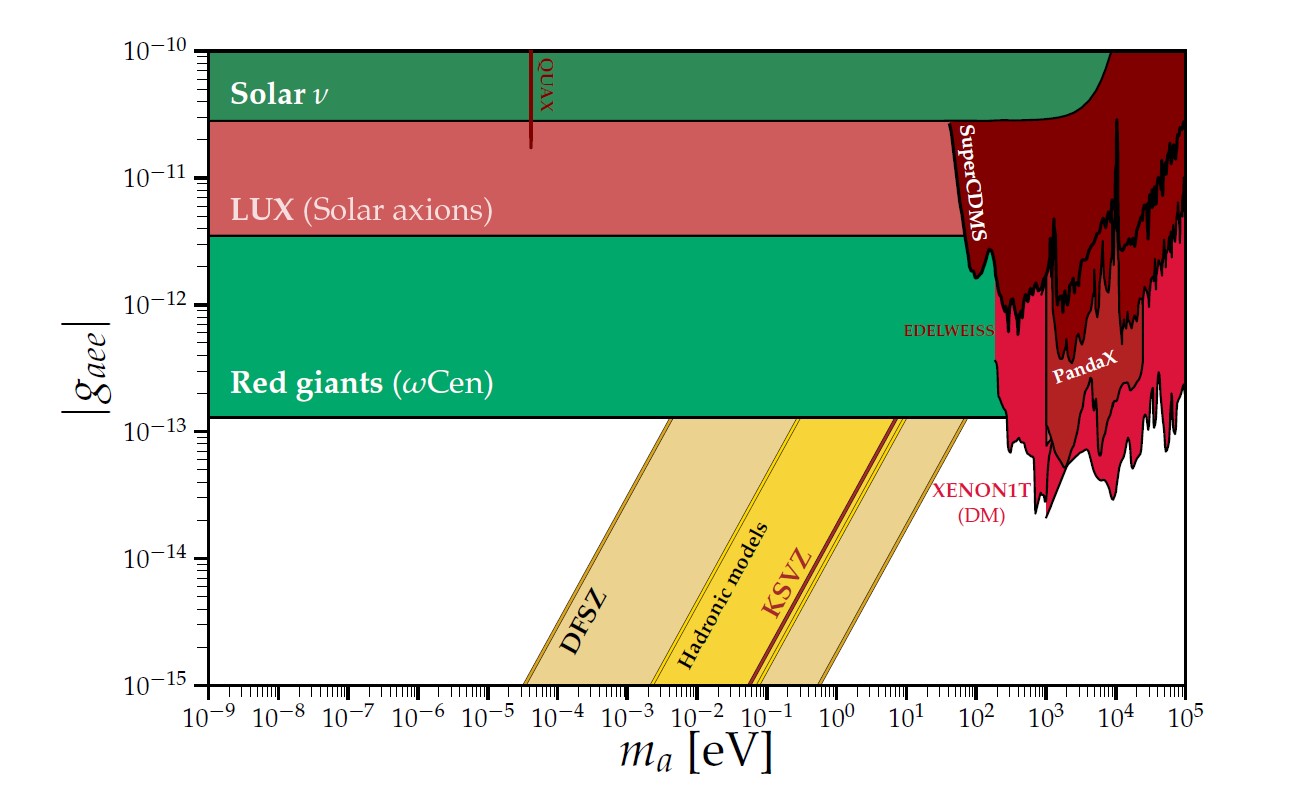}
\caption{\scriptsize Constraints on axion-electron couplings vs mass. (Figure taken from \cite{PDG, gitFigs}.)} \label{Fig:gaeeBound}
\end{center}
\end{figure}

The constraints on matter-axion couplings largely do not come from searches for new axion-mediated macroscopic forces because these turn out to be negligible (at least for a pseudoscalar axion) \cite{Wilczek, GeorgiRandall, Khrip}. But a variety of other observations constrain the effective couplings shown in \pref{AxEFT}, often arising from the absence of too efficient energy loss through axion emission by astrophysical objects like supernovae and red-giant stars. The resulting constraints on $f_a$, $g_{a\gamma\gamma}$ and $g_{aee}$ plotted against the axion mass in Figs.~\ref{Fig:FaxBound}, \ref{Fig:FapBound} and \ref{Fig:gaeeBound} respectively. These plots treat the effective parameters as all being independent, though the QCD axion specifically satisfies the constraint $f_a m_a \simeq \Lambda_\QCD^2$, shown as a diagonal band in the figures. Notice in particular that $f_a \gsim 10^9$ GeV is generally required for most of the displayed mass range. 

\subsubsection{Mechanisms for suppressing vacuum potentials}

We next ask how the QCD picture can be modified so that the vacuum and matter potentials can compete with one another within ordinary matter. We describe here two representative modifications in order to be able to identify their physical consequences.

\subsubsection*{QCD and Yoga suppression}

The most conservative modifications stray the least from the QCD framework, and an example of this type comes from the models of \cite{YogaDE} if the axion in the axio-dilaton pair is imagined to be the QCD axion. The most important feature of these models for the present purposes is that they also contain a relaxation mechanism that is designed to reduce the size of the vacuum energy. Crucially this same mechanism also acts to suppress the size of the vacuum axion potential -- effectively suppressing the scale $\Lambda$ in \pref{QCDVpred} relative to $\Lambda_\QCD$. This suppression makes it easier for potentials inside matter to compete with their vacuum counterparts. But such competition does not require this particular motivation -- see \cite{Hook:2018dlk} for a review of a different mechanism for suppressing the relative size of the vacuum and matter potentials.

\begin{figure}[t]
\begin{center}
\includegraphics[width=120mm,height=80mm]{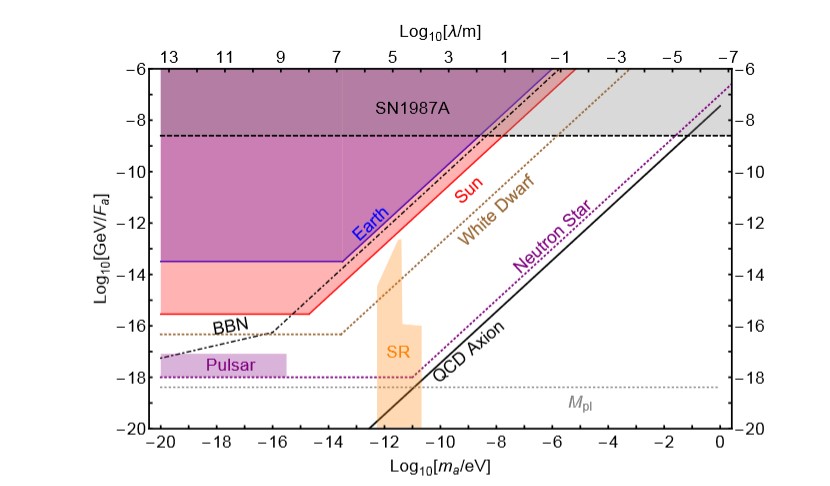}
\caption{\scriptsize Constraint plot showing the excluded region for an axion coupled to QCD but modified to have smaller than usual vacuum potential, showing exclusion regions arising if the axion takes different values inside and outside the Sun, Earth, White Dwarf and Neutron Star. (Figure taken from \cite{Hook:2017psm}.) In hadrophobic models the solar constraints are more relaxed if the axion gradients lie below the photosphere so the zone between the `Earth' and `Sun' curves might then allow laboratory constraints to be satisfied consistent with the screening of the Sun's dilaton charge (on which the strongest solar-system tests depend). More detailed modeling of these constraints is warranted.} \label{Fig:HHBound}
\end{center}
\end{figure}

In this case the axion couples to matter much as does the QCD axion itself and so any variations of the axion profile within macroscopic objects are strongly constrained by measurements of baryonic properties. For instance order-unity deviations of the QCD axion from zero near the surface of the Earth would effectively restore a nonzero QCD vacuum angle and so allow observably large neutron electric dipole moments in laboratory experiments. The observed absence of these phenomena leads to the exclusion curve labeled `Earth' in Fig.~\ref{Fig:HHBound} (taken from \cite{Hook:2017psm}).

Similar arguments constrain the existence of QCD axion excursions within the Sun, leading for instance to changes in the neutron-proton mass difference that would ruin the success of our understanding of solar nuclear reactions. This leads to the exclusion curve marked `Sun' in Fig.~\ref{Fig:HHBound}. Identical reasoning involving the nuclear physics of compact objects like white dwarfs and neutron stars are also shown in the figure. It seems impossible to reconcile well-established properties of nuclear physics with the axion potential needed for screening if the axion involved is a QCD-like axion.

\subsubsection*{Hadrophobic ALP}

The strong constraints on modifying the properties of QCD inevitably suggest hadrophobic constructions in which the model-dependent choices of couplings are used to ensure that axions do not couple directly to QCD -- axion-like particles (ALPS) in particle-physics parlance. To this end the axion symmetry might be designed to act axially on fermions but only for electrons (and possibly other leptons) and not on quarks. The idea is to remove the mixed QCD axion anomaly ($\kappa_s = 0$) and direct couplings to quarks ($C^0_q = 0$). See \cite{Hook:2018dlk, SuperradianceBounds, PDG, gitFigs, AxionMods, Agrawal:2018mkd, Csaki:2019vte, Agrawal:2022lsp, Csaki:2023yas} for variations on axion-matter couplings and their constraints.

In this scenario QCD cannot provide the vacuum potential and so we also introduce a dark nonabelian gauge sector with confinement scale $\Lambda_\ssD$ and dark fermions that transform under the axion symmetry so that the dark sector contributes a mixed axion anomaly (similar to the discussion of \S\ref{sssec:QCDBench}). Being dark, these new particles do not carry colour or electromagnetic charge. Repeating the manipulation leading from \pref{AxEFT0} to \pref{AxEFT} for this new sector then involves performing an axion rotation on both the dark fermions and the SM leptons and so the electrons contribute to $g_{a\gamma\gamma}$ through their contribution to the quantity $\Delta$ and $\kappa_e$. 

In the new basis all nonderivative axion-dependence again arises within mass terms, but this time only for the masses of the dark fermions and SM leptons. Confinement in the dark sector then leads to similar expressions for the vacuum axion potential, with $\Lambda \sim \Lambda_\ssD$ and the role of quark masses being played by any dark-fermion masses. As mentioned in \S\ref{ssec:SimplifiedAxMat}, the appearance of axions in electron masses also then induces a matter-dependent potential inside macroscopic bodies built from nonrelativistic ordinary matter that is proportional to the electron density $n_e(x)$ (and once again the very interactions that cause the axion to acquire gradients inside matter also imply that lepton properties respond to the changed axion values, and so vary with position and time inside -- and on the surface of -- the object).

In this picture the most important traditional constraints on axion-matter couplings are likely those involving $g_{a\gamma\gamma}$ and $g_{aee}$ shown in Figs.~\ref{Fig:FapBound} and \ref{Fig:gaeeBound}. Besides these, there are also new ones coming from the existence of interactions of photons and electrons at the Earth's surface with the assumed nonzero axion profile there. It is unlikely that position-dependent or time-dependent order-unity profiles in $\mfa(x)$ could escape detection on Earth if the axion appears in electromagnetic and/or electron mass interactions like those shown in \pref{AxEFT}, particularly to the extent that the electron mass (and so also atomic energy levels) inherit a spacetime dependence. Such a dependence would require models to lie to the right of the analog of the `Earth' curve of Fig.~\ref{Fig:HHBound} while useful screening in the solar system requires being to the left of the analog of the `Sun' curve. But being to the left of the `Sun' curve is itself only possible if axion excursions under the solar surface are undetectable, which also seems unlikely for axion interactions, particularly if the exterior minimum is a maximum of the matter component of the potential -- as is the case for the potential \pref{QCDmattax} for example -- as argued in \cite{Hook:2017psm}. We return to some of these bounds in \S\ref{ssec:SimplifiedAxMat}.

Although the above worries are less important for applications to screening outside of the solar system, finding a formulation of axion-matter couplings that can both generate nontrivial minima inside bulk matter and be consistent with particle phenomenology remains a weak link for any realistic applications of this type of screening within the solar system. This motivates a more careful enumeration of the constraints themselves, in addition to ways in which they can be evaded. One approach to model-build around the problem might be to couple the axion to the Dark Matter density, and work in a regime where appreciable amounts of Dark Matter accrete and accumulate within the Sun.

\subsection{Simplified axion-matter couplings}
\label{ssec:SimplifiedAxMat}

So is it hopeless? Can acceptable axion-matter couplings be constructed without running into observational problems? In this section we step back and explore the possibility of general axion couplings to matter without the more detailed predictions coming from the lagrangian \pref{AxEFT0} (along the lines also explored in \cite{Olive:2007aj, Hees:2018fpg, Banerjee:2022sqg, Beadle:2023flm}). We continue to assume the existence of a vacuum axion potential $V(\mfa)$ and imagine it to be of size $\Lambda_\ssD^4 = m_{\rm out}^2 f_a^2$ times an arbitrary dimensionless $\cO(1)$ function of $\mfa$ that is minimized at $\mfa_+$. 

We also seek a matter coupling that can both generate a matter-dependent axion potential and not already be ruled out experimentally, and to this end couple $\mfa$ to fermionic fields $\psi$ through a phenomenological interaction of the form
\be\label{PhenoUePot}
    \cL_{\mfa m}= - \sqrt{-g} \; U(\mfa) \, \ol \psi \psi \,.
\ee
We assume $U(\mfa)$ is order some scale $\Lambda_m$ times a dimensionless $\cO(1)$ function of $\mfa$ that is minimized at $\mfa=\mfa_- \neq \mfa_+$. This coupling generates a bulk axion potential within matter because $U(\mfa)$ enters into the energy of bulk matter in the same way as would an axion-dependent electron mass. For macroscopic nonrelativistic systems the corresponding energy density therefore is 
\be
  \delta\cE \simeq U(\mfa) n(x) \simeq \frac{U(\mfa)}{m_\ssN} \, \rho(x) \,, 
\ee
where $n(x)$ denotes the relevant fermion density (in practice electron or baryon number), while $\rho(x)$ is the local mass density and $m_\ssN$ is the nucleon mass. 

\begin{figure}[h]
\begin{center}
\includegraphics[width=70mm,height=40mm]{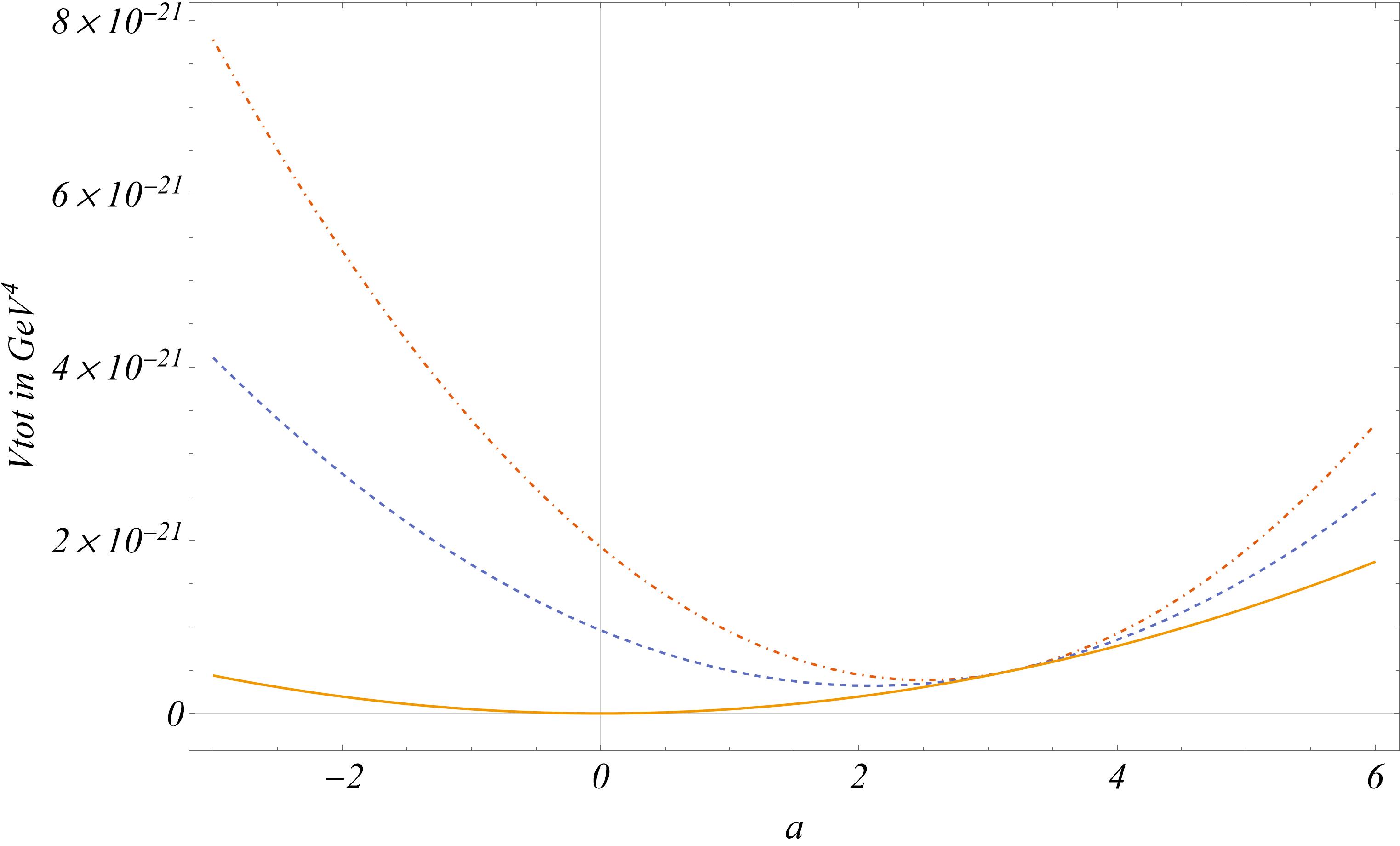}
\includegraphics[width=70mm,height=40mm]{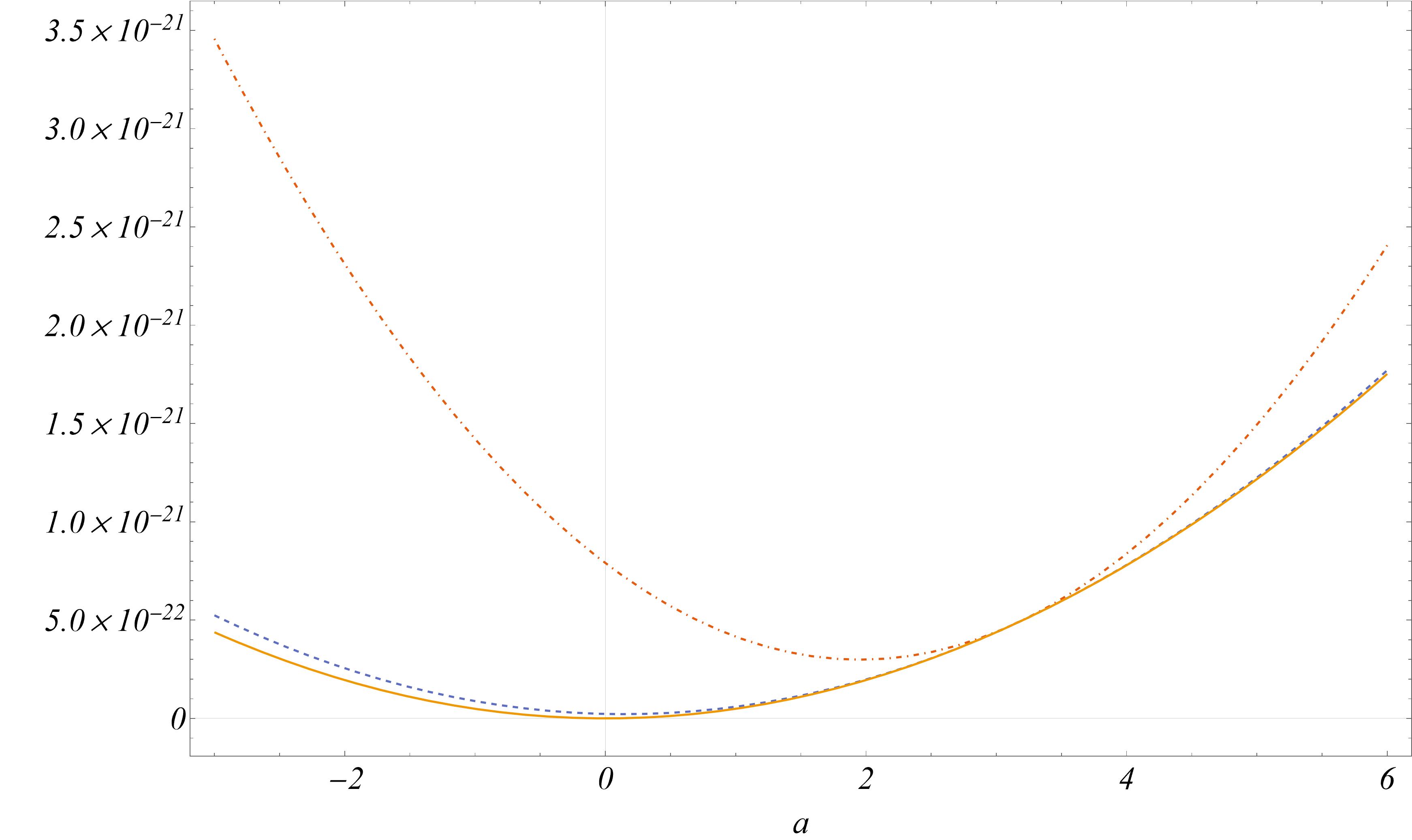}
\caption{\scriptsize Vacuum plus matter potential evaluated at specific radii, for the two density profiles showin in Fig.~\ref{Fig:nvsrPlots}. Left panel (step function profile) evaluates at $r = 0.5 R$ (dot-dashed), $R$ (dotted) and $2R$ (solid). Right panel (exponential profile) evaluates at $r = 0.1 R$ (dash-dotted), $0.5R$ (dotted) and $2R$ (solid). } \label{Fig:VtotvsrPlots}
\end{center}
\end{figure}

The interaction of \pref{PhenoUePot} is not so different from the axion discussion of the previous section except that we do not require $U(\mfa)$ to be the specific trigonometric function predicted by an axial rotation -- so $U(\mfa)$ need not be an order-unity axion-dependence in the fermion mass, for example. In practice we imagine both $V(\mfa)$ and $U(\mfa)$ to be near their local minima and so take $V(\mfa) \propto (\mfa - \mfa_+)^2$ and $U(\mfa) \propto (\mfa - \mfa_-)^2$, leading to position-dependent potentials of the form shown in Fig.~\ref{Fig:VtotvsrPlots} once combined with the two choices of density profile shown in Fig.~\ref{Fig:nvsrPlots}. In the absence of tuning both the position of the minimum and the value of the potential at the minimum generically change in a position dependent way.

The axion's behaviour is controlled by the parameters $f$, $\Lambda_\ssD$ and $\Lambda_m$, and we choose these using the following criteria:
\begin{enumerate}
\item We choose the matter scale $\Lambda_m$ to ensure the matter interaction \pref{PhenoUePot} evades the constraints described in the previous section. To this end we conservatively ask that $U(\mfa)$ not shift particle masses $\delta m_\psi/m_\psi \simeq U(\mfa)$ by more than a part in $10^{15}$ so as to not contribute observably to atomic clock measurements on earth (see \cite{Sherrill:2023zah} for a recent analysis of the sensitivity of these measurements).\footnote{This condition also ensures that other constraints -- such as the contribution of the axion gradient to the internal pressure and so on -- are also automatically satisfied.} This requires $\Lambda_m \lsim 10^{-15} \, m_\ssN \sim 10^{-6}$ eV. Such small values also ensure that matter loops contribute negligibly to the axion mass.
\item We choose the decay constant $f$ so that the axion mass internal to a macroscopic body is large enough to justify the adiabatic approximation ({\it i.e.}~we choose its Compton wavelength to be much smaller than the scale of density variations within the source). For instance, using \pref{EquivVacScale} together with the values of Table \ref{TableMacAxionMass} shows that asking $m_{\rm in}^{-1} \lsim 200$ km requires $m_{\rm in} \gsim 10^{-12}$ eV and so $f \lsim 10^3$ GeV. This is at face value ruled out by the constraints shown in Fig.~\ref{Fig:FaxBound}, a point to which we return and re-examine below.
\item We choose $\Lambda_\ssD$ so that the axion mass outside of the source is systematically small relative to its value inside, in order to localize the axion gradient to lie near the object's surface (see Fig.~\ref{Fig:axvsrPlots} and the associated discussion surrounding it) and so justify the narrow-width analysis. For instance, choosing $m_{\rm out}/m_{\rm in} \sim 10^{-3}$ implies $m_{\rm out} \sim 10^{-16}$ eV (and so $m_{\rm out}^{-1} \sim 2\times 10^5$ km) if we choose $m_{\rm in}^{-1} \sim 200$ km.
\end{enumerate}

\begin{figure}[h]
\begin{center}
\includegraphics[width=70mm,height=40mm]{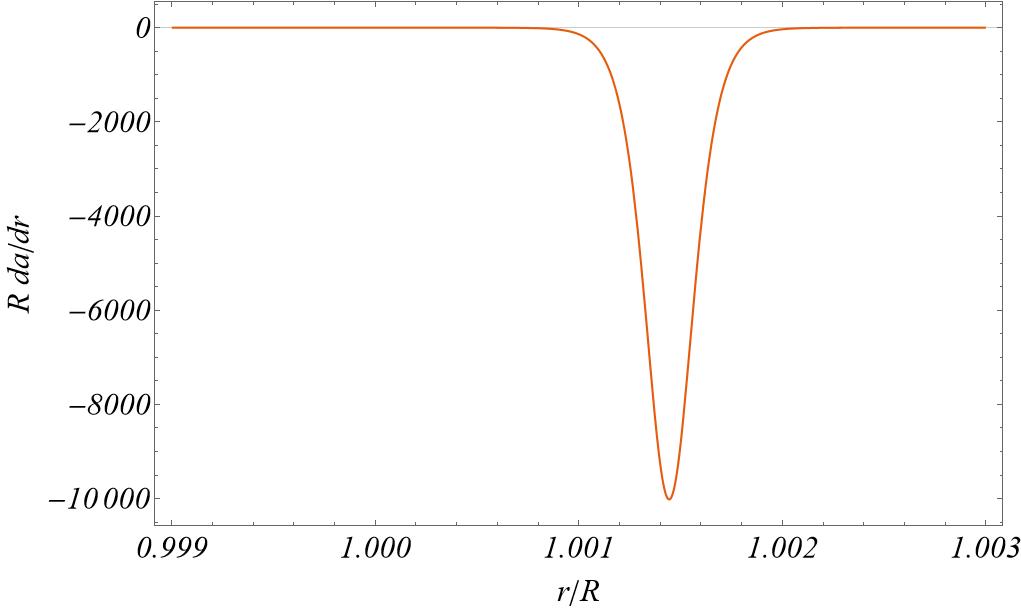}
\includegraphics[width=70mm,height=40mm]{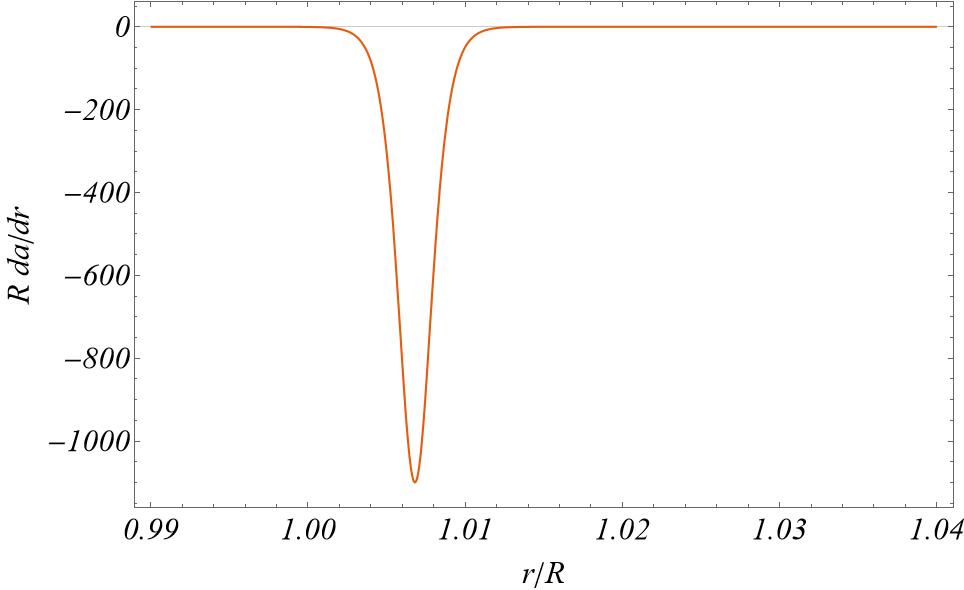}
\caption{\scriptsize Calculated axion derivative $R (\exd \mfa/\exd r)$ as a function of radius, for the two density profiles shown in Fig.~\ref{Fig:nvsrPlots}. Left panel uses the step function profile and right panel uses the exponential profile. These profiles assume the axion mass is high enough that the axion adiabatically follows the local minimum, and as a result the profile variation is set by the variation of source properties. } \label{Fig:dphidrvsrPlots}
\end{center}
\end{figure}

The axion derivative computed using the assumptions $m_{\rm in}^{-1} \sim 200$ km, $m_{\rm out}/m_{\rm in} \sim 10^{-3}$ and $f \sim 10^3$ GeV is shown in Fig.~\ref{Fig:dphidrvsrPlots}, verifying that the adiabatic and narrow-width approximations both apply for both the step-function and exponential density profiles, with the gradient being appreciable in a region of size $\ell \sim 0.0003 R$ for the step distribution and $\ell \sim 0.003 R$ for the exponential profile, leading to an acceptable level of screening of strength $\mfg_{\oplus}/\mfg \sim 10^{-4}$ for the step profile (Earth) and $\mfg_{\odot}/\mfg \sim 10^{-3}$ for the exponential profile (Sun).

What about the constraints of Fig.~\ref{Fig:FaxBound} that seem to rule out having $f$ as low as $10^3$ GeV? Significantly, these constraints come from energy-loss bounds that place upper limits on the efficiency with which astrophysical objects can drain their energy through axion emission.\footnote{Axions with small decay constants need not freely stream, see \cite{Lella:2023bfb} for a recent treatment of bounds in this case.} This means that they constrain the linear coupling of the axion fluctuation to the matter in these objects. The bounds are derived under the assumption that this coupling is equally large throughout the entire object, which is not true in the present case because this coupling turns off when $\mfa$ sits near the minimum of $U(\mfa)$ (as it does everywhere except for a strip of width $\ell$ near the object's surface.

\begin{figure}[h]
\begin{center} 
\includegraphics[width=80mm,height=60mm]{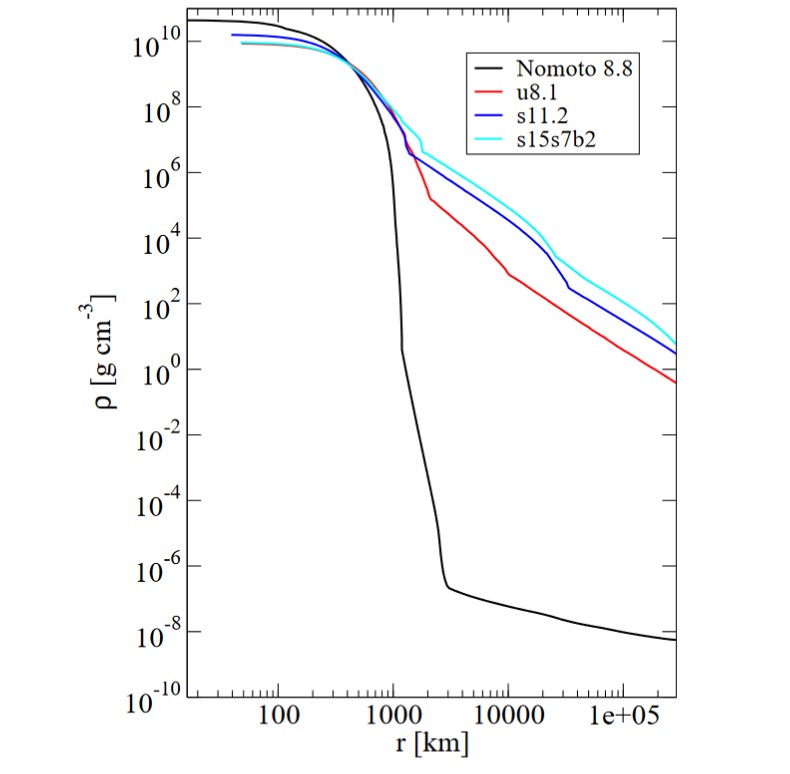}
\caption{\scriptsize Core-density profiles of different supernova progenitors at the onset of gravitational collapse with different curves corresponding to different assumptions about progenitor properties, as described in detail in \cite{Janka:2012wk}.} \label{Fig:SNPlot}
\end{center}
\end{figure}

In principle these bounds should be re-evaluated from scratch but an estimate for what might be expected can be found along the lines considered\footnote{Ref.~\cite{Olive:2007aj} consider the constraints due to energy loss through an $\mfa^2 \ol\psi \psi$ coupling and so find an emission rate that goes as $1/f^4$ (and so is weaker than the linear emission from the boundary region of significant gradient that we consider here \cite{Banerjee:2022sqg}).} in \cite{Olive:2007aj, Banerjee:2022sqg}. The local emission rate is proportional to $\rho_\ssB^2 T^{1/2}$ \cite{Chang:2018rso} where $\rho_\ssB$ is the local baryon mass density and $T$ is the medium's local temperature. The bound $f \gsim 10^9$ GeV of Table \ref{Fig:FaxBound} might plausibly be replaced by $F_{\rm eff} \gsim 10^9$ GeV where 
\be
   \frac{1}{F_{\rm eff}^{2}} \sim \left( \frac{\rho_s}{\ol\rho} \right)^{2} \left( \frac{T_s}{\ol{T}} \right)^{1/2} \left( \frac{\ell}{R} \right) \frac{1}{f^2} \,,
\ee
where $\ol\rho$ and $\ol{T}$ are the characteristic density and temperature relevant to the standard calculation and $\rho_s$ and $T_s$ are their counterparts in the region of nonzero axion gradient near the surface. The factor $\ell/R \sim 10^{-3}$ accounts for the reduced volume of the region in which linear axion-matter couplings are significant (with the numerical estimate assuming these are similar in size as found earlier for the Sun). The bound $F_{\rm eff} \gsim 10^9$ GeV would be consistent with $f \sim 10^3$ GeV provided $(\rho_s/\ol{\rho})^2(T_s/\ol{T})^{1/2} \lsim 10^{-9}$. For example, if $\rho \sim m_\ssN(m_\ssN T)^{3/2}$ were described by a nonrelativistic thermal density then $(T_s/\ol{T})^{7/2} \lsim 10^{-9}$ would require $T_s/\ol{T} \lsim 0.003$. This seems to be a relatively mild condition based on steeply falling density profiles obtained from stellar and supernova simulations (see for instance Fig.~\ref{Fig:SNPlot} \cite{Janka:2012wk}). It may well be that $f \sim 10^3$ GeV remains viable though more detailed calculation are required to be sure.

A decay constant $f \sim 10^3$ GeV also evades other constraints on particle scattering, such as the non-observation of axion exchange or emission at high-energy particle accelerators. It seems that the parameter regime to which we are led merits more careful exploration.

\section{Conclusions}
\label{sec:Conclusions}

Gravitationally coupled light scalar fields can appear in the low-energy limit of fundamental physics, often as pseudo-Goldstone bosons for underlying accidental internal and scaling symmetries. They often play an important role in proposals to address the dark energy problem. They may have a range of experimental implications that are amenable to current and future experimental tests.

Single-field models are the simplest to consider and have been the most explored, including in particular the axion and dilaton (Brans-Dicke) scalars suggested by low-energy shift symmetries. But single-field models are not representative of the general low-energy properties of light scalar fields because they are not complicated enough to allow the two-derivative self-interactions that most naturally compete with the two-derivative metric interactions of General Relativity. Furthermore, in string and supergravity theories dilaton and axion fields arise together as the real and imaginary components of complex scalars. The pseudoscalar nature of axions often makes them less constrained by long-range tests of gravity but their kinetic couplings to dilatons make axio-dilatons natural generalization to simple Brans Dicke theories.

In this article we consider some phenomenological aspects of axio-dilaton systems that are light enough to be relevant to long-distance tests of gravity. In particular we examine the strong experimental solar-system constraints that restrict dilaton couplings to be at most of order $\mfg^2 \lsim 10^{-5}$ as compared to gravitational couplings, and ask whether axio-dilaton self-interactions can allow these bounds to be relaxed. The great potential importance of these fields, both theoretically and experimentally, suggests exploring ways to evade these bounds, such as through `screening' mechanisms in which nonlinearities of interactions make the field generated by a macroscopic body smaller than the sum of the fields generated by its constituent particles. Several such mechanisms are known for single-field extensions of GR \cite{Khoury:2003aq, Hinterbichler:2010es}, that can sometimes successfully evade experimental constraints (and might be subject to test in the near future).

In \cite{Homeopathy} it was remarked that two-derivative axio-dilaton scalar interactions offer great potential for providing new screening mechanisms, because very small axion gradients can dramatically modify the dilaton field surrounding a source. Unfortunately more detailed modelling of linear axion-matter interactions within sources show that the simplest sources do not exploit this mechanism once the solutions outside a source(like the Sun or Earth) are matched onto those inside a source \cite{Brax:2022vlf, Lacombe:2023qfx}. In this article we consider a closely related alternative formulation of the axion-dilaton coupling for which screening of the dilaton does appear to be possible. We do so by postulating axion matter interactions that give rise to axion gradients, which in turn modify the induced dilaton profile through the kinetic axion-dilaton coupling $W(\phi)$.

For some choices for $W(\phi)$ this reduction can give a suppression of a macroscopic body's dilaton coupling by the ratio of the width of the region containing the axion excursion and the source radius, similar to what is achieved in Chameleon models. This can lead to the required suppression of $\mfg$ by the required factor of $10^{-3}$. We call this the Axio-Chameleon mechanism to highlight the similarity and differences with the Chameleon model.  In our mechanism suppression is achieved as the integration constants adjust themselves to minimize the energy of the system, though at the price of introducing frustration in the scalar field in the presence of multiple sources. We argue that the field resolves this frustration by adjusting to dominantly suppress the dilaton charge of the source with the largest surface gravity, which is the Sun in the solar system.

The main hurdle these models face are particle-physics constraints on the required axion couplings, which can lead to particle properties varying with the axion gradients. We are able to evade these limits but only at the expense of having axion decay constants as low as $f \sim 10^3$ GeV, that are much lower than those normally considered. Although this naively runs up against energy-loss constraints from astrophysical objects these constraints must be re-evaluated given that the linear axion-matter coupling tends to shut itself off wherever the axion gradient is not important \cite{Olive:2007aj, Banerjee:2022sqg}. This suppresses the energy-loss rate by a geometrical factor $(\ell/R)$ because axion gradients are localized near an object's surface, and by a factor that captures how cool the surface is relative to its interior.  

We hope our proposal sparks further studies of screening (and general gravitational response) for multiple scalars interacting through two-derivative sigma-model couplings, since these are both well-motivated and relatively poorly explored, for which the competition with GR at low energies is likely to contain a rich variety of new low-energy phenomena.

\section*{Acknowledgements}
We thank Tessa Baker, Sebastian Ellis, Kurt Hinterbichler, Maria Mylova, Johannes Noller and Junwu Huang for helpful conversations. CB thanks the Benasque Center for Physics and CERN for providing the pleasant environs in which some of these ideas were developed. CB's research was partially supported by funds from the Natural Sciences and Engineering Research Council (NSERC) of Canada. Research at the Perimeter Institute is supported in part by the Government of Canada through NSERC and by the Province of Ontario through MRI. The work of FQ has been partially supported by STFC consolidated grants ST/P000681/1, ST/T000694/1.

\end{document}